\documentclass[sigconf]{acmart}

\title{Replay Clocks}
\author{Ishaan Lagwankar}
\affiliation{
    \institution{Michigan State University}
    \country{East Lansing, Michigan, USA}
}
\email{lagwanka@msu.edu}
\author{Sandeep S Kulkarni}
\affiliation{
    \institution{Michigan State University}
    \country{East Lansing, Michigan, USA}
}
\email{sandeep@cse.msu.edu}
\date{May 2023}

\renewcommand\footnotetextcopyrightpermission[1]{}

\usepackage{graphicx} 
\usepackage{algorithm}
\usepackage[noend]{algpseudocode}
\usepackage{xspace}
\usepackage{color}
 \usepackage{hyperref} 
 \usepackage{adjustbox}
\usepackage{natbib}
\usepackage[belowskip=-8pt,aboveskip=10pt]{caption}
\usepackage{subcaption} 
\newtheorem{assumption}{Assumption}

\newcommand{\br}[1]{\ensuremath{\langle #1 \rangle}\xspace}

\edef\qedrestoreat{\noexpand\catcode\lq\noexpand\@=\the\catcode\lq\@}
\catcode\lq\@=11

\ifx\protect\undefined\let\protect\relax\fi

\def\qed{\protect\@qed{$\qedsymbol$}}

\def\QED{\protect\@qed{{\rm Q.E.D.}}}

\def\QEI{\protect\@qed{{\rm Q.E.I.}}}

\def\QEF{\protect\@qed{{\rm Q.E.F.}}}

\def\Proof{\protect\@Proof}\def\endProof{\protect\@endProof}%

\def\qedsymbol{\square}

\def\TheWordProof{\bf Proof\enskip}

\def\ProofFont{}

\newif\ifAutoQED\AutoQEDfalse

\newif\ifNumberResults
\ifx\theorem@style\undefined
   \NumberResultstrue
   \def\TheoremHeader#1#2#3{\bf #1\ifNumberResults\ #2\unskip\fi#3}
   \def\TheoremFont{\it}
\fi

\def\TheoremsAsCommands{%
  \def\TheoremFont{}
  \def\begin@theorem##1{%
     \par\removelastskip\smallskip
     \save@set@qed 
     \noindent{##1}
   }%
  \def\@endtheorem{\ifAutoQED\qed\fi\restore@qed}%
}%

\expandafter\let\csname ds@theorems-as-commands\endcsname\TheoremsAsCommands

\def\parag@pushright#1{{
    \parfillskip=0pt            
    \widowpenalty=10000         
    \displaywidowpenalty=10000  
    \finalhyphendemerits=0      
    \hbox@pushright             
    #1
    \par}}

\def\hbox@pushright{
    \unskip                     
    \nobreak                    
    \hfil                       
    \penalty50                  
    \hskip.2em                  
    \null                       
    \hfill                      
}%

\def\vbox@pushright#1{\expandafter\message 
  {QED.sty could be improved in this case (line \the\inputlineno): please ask}%
  \page@pushright{#1}}%

\newif\if@qed\@qedfalse
\def\save@set@qed{\check@pt@sty@v\let\saved@ifqed\if@qed\global\@qedtrue}%
\def\restore@qed{\global\let\if@qed\saved@ifqed}

\def\@Proof{%
   \par\removelastskip\smallskip\penalty700
   \save@set@qed
   \noindent\ProofFont{\TheWordProof\enskip}%
}%

\def\@endProof{%
   \qed\restore@qed
   \penalty-700 \smallskip
}

\def\@qed#1{\check@pt@fm@thm
\if@qed                                 
     \global\@qedfalse\pushright{#1}
\else\ifhmode\ifinner\else\par\fi\fi
\fi}

\def\@pushright#1{%
  \ifvmode
       \ifinner\vbox@pushright{#1}
       \else   \page@pushright{#1}%
       \fi
  \else\ifmmode\maths@pushright{\hbox{#1}}
       \else\ifinner\hbox@pushright{#1}
            \else\parag@pushright{#1}
  \fi  \fi  \fi
}

\def\maths@pushright#1{%
  \ifinner
     \hbox@pushright{#1}%
  \else
     \eqno#1
     \def\]{$$\ignorespaces}
  \fi
}

\def\page@pushright#1{
  \skip@\lastskip
  \ifdim\skip@>\z@
       \unskip    
  \else\skip@\z@\relax
  \fi
  \dimen@\baselineskip
  \advance\dimen@-\prevdepth            
  \nobreak                              
  \nointerlineskip
  \hbox to\hsize{%
    \setbox\z@\null
    \ifdim\dimen@>\z@\ht\z@\dimen@\fi   
    \box\z@
    \hfill
    #1}%
  \vskip\skip@                  
}%

\ifx\nonqed@thm\undefined 
   \let\nonqed@thm\@thm 
   \let\nonqed@endthm\@endtheorem
\fi

\def\@thm{\save@set@qed\nonqed@thm}%
\def\@endtheorem{\ifAutoQED\qed\fi\restore@qed\nonqed@endthm}%

\newbox\qed@box 

\def\WillHandleQED{\relax
\ifx\HandleQED\nohandle@qed
   \def\pushright{\global\setbox\qed@box\hbox}
   \let\QEDbox\qed@box 
   \def\HandleQED{\unhbox\QEDbox}
   \aftergroup\check@handle@qed  
\else
   \let\QEDbox\voidb@x 
\fi}

\def\nohandle@qed{%
\errhelp{One of them is missing: see QED.sty.}%
\errmessage{This environment uses \string\WillHandleQED\space and
\string\HandleQED\space incorrectly}}

\def\check@handle@qed{\relax
\ifvoid\qed@box\else\expandafter\nohandle@qed\fi}

\def\UnHandleQED{%
\let\HandleQED\nohandle@qed
\let\QEDbox\voidb@x
\def\pushright{\protect\@pushright}}%

\UnHandleQED

\def\obsolete@pt@version#1{\errhelp={%
        Anonymous FTP ftp.dcs.qmw.ac.uk /pub/tex/contrib/pt/tex/#1}
\errmessage{You have an obsolete version of #1 - please get a new one}}%

\ifx\theorem@style\undefined
   \def\check@pt@fm@thm{\relax
     \ifx\square\undefined
       \gdef\square{\bigcirc
          \errhelp={Anonymous ftp e-math.ams.com /ams/amsfonts}%
          \errmessage{\string\square\space is an AMS symbol}%
          \global\let\square\bigcirc}%
     \fi
      \ifx\theorem@style\undefined
          \global\let\check@pt@fm@thm\relax
      \else\errhelp={The macros \@thm and \@endtheorem need to be re-defined.}%
          \errmessage{QED.sty must be loaded AFTER theorem.sty but before
           using \string\newtheorem}%
      \fi
      \global\let\check@pt@fm@thm\relax
      }%
\else
   \def\check@pt@fm@thm{%
      \ifx\square\undefined
         \def\square{\bigcirc
         \errhelp={Anonymous ftp e-math.ams.com /ams/amsfonts}%
         \errmessage{\string\square\space is an AMS symbol}%
         \global\let\square\bigcirc}%
      \fi
      \global\let\check@pt@fm@thm\relax
      }%
\fi

\def\check@pt@sty@v{%
   \relax
   \ifx\@@begintheorem\undefined
   \else\message{*** QED.sty and the old Paul.sty seriously conflict! ***}%
        \obsolete@pt@version{Paul.sty}%
   \fi
   \ifx\execHorizontalMap\undefined
        \gdef\obsolete@pt@version##1{}
   \else\global\let\di@gr@m\diagram
        \gdef\diagram{\obsolete@pt@version{diagrams.tex}\di@gr@m}%
   \fi
   \global\let\check@pt@sty@v\relax
}%

\def\DefineStandardTheorems{%
\relax\ifx\@@thm\undefined
        \@addtoreset{Result}{section}%
        \ifx\chapter\undefined\else\@addtoreset{Result}{chapter}\fi
\fi
\let\theUnnumbered\relax\countdef\c@Unnumbered255 \def\p@Unnumbered{}%
\let\DefineStandardTheorems\relax
}

\ifx\newtheorem\undefined\else\ifx
\else
\fi\fi

\ifx\theorem@style\undefined\else\qedrestoreat\expandafter \fi

\def\@opargbegintheorem#1#2#3{\begin@theorem{\TheoremHeader{#1}{#2}{ (#3)} }}

\def\@begintheorem#1#2{\begin@theorem{\TheoremHeader{#1}{#2}{} }}

\def\begin@theorem#1{%
  \trivlist\item[\hskip\labelsep {#1}]
  \TheoremFont                                  
  \ifx\ProofFont\empty\def\ProofFont{\rm}\fi    
  }%

\ifx\newtheorem\undefined\else\qedrestoreat\expandafter \fi

\def\ther@sult#1{%
\ifx\chapternumber\undefined\else
   \ifnum\chapternumber>0 \number\chapternumber.\fi\fi
\ifx\sectionnumber\undefined\else
   \ifnum\sectionnumber>0 \number\sectionnumber.\fi\fi
\ifx\subsectionnumber\undefined\else
   \ifnum\subsectionnumber>0 \number\subsectionnumber.\fi\fi
\expandafter\number\csname c@#1\endcsname}%

\def\plain@thm#1#2{
  \save@set@qed
  \expandafter\advance\csname c@#1\endcsname1
  \@begintheorem{#2}{\csname the#1\endcsname}%
}%

\def\newtheorem#1[#2]#3{
  \expandafter\def\csname #1\endcsname{\plain@thm{#2}{#3}}%
  \expandafter\def\csname end#1\endcsname{\@endtheorem}%
  \expandafter\ifx\csname c@#2\endcsname\relax
     \expandafter\expandafter\csname newcount\endcsname
        \csname c@#2\endcsname
     \expandafter\def\csname the#2\endcsname{\ther@sult{#2}}%
  \fi
}%

\TheoremsAsCommands

\qedrestoreat 

\newtheorem{theorem}{\textbf{Theorem}}

\newtheorem{observation}{\textbf{Observation}}
\newtheorem{lemma}{\textbf{Lemma}}
\newcommand{\ignore}[1]{}
\newif\ifspace\spacefalse
\newif\ifspacetwo\spacetwofalse

\usepackage{amsmath}
\usepackage{amsthm}
\theoremstyle{acmplain}

\newcommand{\newhvc}{RepCl\xspace}

\newcommand{\observe}{\vspace*{1mm} \noindent {\bf Observation}\xspace}
\newcommand{\newdefine}{\vspace*{1mm} \noindent {\bf Definition}\xspace}
\newcommand{\newlem}{\vspace*{1mm} \noindent {\bf Lemma}\xspace}

\newcommand{\ShiftEpochComplexity}{1\xspace}
\newcommand{\SendReceiveComplexity}{2\xspace}
\newcommand{\Comparetimestamps}{3\xspace}

\newcommand{\newhvcless}{1\xspace}
\newcommand{\concurtimestamps}{2\xspace}

\newcommand{\newhvchappenedbefore}{1\xspace}
\newcommand{\newhvcconcur}{2\xspace}
\newcommand{\effar}{3\xspace}
\newcommand{\allorder}{4\xspace}

\newcommand{\epsilonone}{\Bigepsilon_1}
\newcommand{\epsilontwo}{\Bigepsilon_2}
\newcommand{\maxt}{mx\xspace}
\newcommand{\newmaxt}{newmx\xspace}

\newcommand{\maxph}{mpt\xspace}

\newcommand{\hb}{\xspace\textsf{\xspace hb}\xspace}

\newcommand{\intervalsize}{\ensuremath{I}\xspace}

\newcommand{\Bigepsilon}{\mathcal{E}\xspace}

\newcommand{\offsetsize}{\tau\xspace}
\newcommand{\countersize}{\sigma\xspace}

\begin{document}

\begin{abstract}

In this work, we focus on the problem of replay clocks ($\newhvc$). The need for replay clocks arises from the observation that analyzing distributed computation for all desired properties of interest may not be feasible in an online environment. 
These properties can be analyzed by replaying the computation. However, to be beneficial, such replay must account for all the uncertainty that is possible in a distributed computation. Specifically, if event $e$ must occur before $f$ then the replay clock must ensure that $e$ is replayed before $f$. On the other hand, if $e$ and $f$ could occur in any order then replay should not force an order between them. 

After identifying the limitations of existing clocks to provide the replay primitive, we present $\newhvc$ and identify an efficient representation for the same. We demonstrate that $\newhvc$ can be implemented with less than four integers for 64 processes for various system parameters if clocks are synchronized within $1ms$.
Furthermore, the overhead of $\newhvc$ (for computing/comparing timestamps and message size) is proportional to the size of the clock. Using simulations, we identify the expected overhead of $\newhvc$ based on the given system settings. We also identify how a user can the identify feasibility region for $\newhvc$. Specifically, given the desired overhead of $\newhvc$, it identifies the region where unabridged replay is possible.  

\end{abstract}

\maketitle

\section{Introduction}

According to the observer effect, when we try to measure something, we change it to some extent. 
Therefore, precise measurement is never truly possible. Computer programs suffer from this same difficulty. When you try to measure something in a computation, it changes the underlying computation. In an ideal world, a program may want to make sure that every step that it is taking is correct with respect to any environmental changes. However, the time taken for performing these checks may cause the program to be incorrect. In other words, it is possible that adding excessive safety checks or checks for guaranteeing fairness may cause the system to spend substantial time on those checks and thereby violate system requirements even though those requirements would never have been violated without those checks in the first place. 

This issue is even more complicated in distributed computing where each process (component, node, etc.) relies on partial information. Hence, computing the required safety checks (or checking for the satisfaction of fairness requirements, etc.) would require processes to communicate with each other. In turn, the time for computing them would be even higher. 

As an illustration, consider two drones $A$ and $B$ that are cooperating to perform a task. Each drone may take independent actions based on some environment that the other drone cannot see. It is required that the area covered by the drones remains 50\% or above at all times (100\% of the time). It is also preferred that this area remains at 75\% most of the time (75\% of the time). 
Here, we would like to know (1) how frequently a given point $x$ was covered by one of the drones, (2) how frequently a given point was covered by both the drones, (3) what was the minimum coverage at any time, etc. 
(Assume that they are at two different altitudes so safety measures such as preventing collision are not necessary). 
Doing all these checks during the execution would require that $A$ and $B$ communicate with each other before they make any move. In turn, it would change the behavior of the drones completely thereby preventing us to make any conclusion about their behavior in the absence of these checks. Furthermore, the problem would be even more complicated if we had a larger number of drones. 

One way to address this problem is to log the computation as it is happening so we can evaluate it later for all properties of interest.
These properties may include non-critical safety properties or desirable performance criteria, etc. To be beneficial, the amount of storage for the log or the time to create that log should be small enough so that the underlying computation is affected minimally.
At the same time, the log should capture the non-determinism that is inherently present in any distributed computation.  
We also want to make sure that the creation of the logs is performed independently by each process, i.e., each process stores its local state whenever it changes along with a timestamp (discussed next) that identifies when the change was made. 
We consider various approaches for storing the timestamps and their implications.  

The simplest approach we can consider is to let the timestamp be the physical time of the relevant process. Here, the storage and computation cost is very low. However, the physical clocks of processes often differ. Hence, it is possible that drone $A$ may send a message at time 50 (local time of $A$) but it is received by $B$ at time $40$ ($B$'s local time). When we try to reply to this log to evaluate the given properties, it will cause $B$ to receive the message before $A$ has sent it. This is unacceptable. 

The next approach we can consider is vector clocks. Vector clocks introduce two concerns: Their size of $O(n)$, where $n$ is the number of processes, may be too high. Another challenge is that vector clocks do not have any reference to the physical clock and do not account for communication outside the system. For example, it is possible that drone $A$ activated a green LED at time $t_1$ and $B$ activated a white LED at time $t_2$ where $t_2 >> t_1$. In other words, an external observer will know that the action of $A$ occurred before $B$. However, if $A$ and $B$ did not communicate then the corresponding events will be concurrent \cite{Lamport78CACM}. Thus, when we replay the log, it is possible that the white LED event could be replayed before the green LED event. This is also unacceptable. 

Hybrid logical clocks (HLC) \cite{kulkarni2022}\cite{kulkarni2014logical} combine logical clocks and physical clocks. Specifically, they rely on a system where physical clocks are synchronized within the acceptable limit of clock skew, $\Bigepsilon$, they guarantee that $hlc.e < hlc.f$ if $e$ happened before $f$ or $pt.e+\Bigepsilon < pt.f $ (\cite{Lamport78CACM}). Here, $hlc.e$ denotes the Hybrid Logical Clock of process $e$, and $pt.e$ denotes the physical time observed on process $e$. In other words, $hlc.e < hlc.f$ if $f$ causally depends upon $e$ or $f$ occurred substantially after $e$. 
They eliminate the problem associated with physical clocks as HLC respects causality. They also eliminate the problem caused by vector clocks as the $HLC$ timestamp of activating the green LED will be less than the $HLC$ timestamp of activating the white LED. 
$HLC$ does create another problem though. Consider the case where we have events $e$ and $f$ such that $|pt.e - pt.f| < \Bigepsilon$ and $e||f$, i.e., the events are causally concurrent and very close to each other in physical time.  Without loss of generality, let $hlc.e < hlc.f$. In this situation, when we replay the log, $e$ will always occur before $f$. In other words, the log does not have the necessary information that could allow it to replay $f$ before $e$ even though they could have occurred in any order.
An extension of HLC, hybrid vector clocks \cite{yingchareonthawornchai2018analysis} reduces some of these issues. However, as we highlight in Section \ref{sec:properties}, this overhead is still quite high. 

Based on these limitations, in this paper, we focus on building a new clock, \newhvc, that combines hybrid logical clocks and vector clocks so as to eliminate their limitations. Our goal is to investigate scenarios under which \newhvc permits efficient replay of events. To understand why we may need to limit \newhvc to specific scenarios, observe that if the underlying system was asynchronous (unbounded clock drift) then it is required to have $O(n)$ vector clocks to enable replay of events. Systems that communicate frequently will need more information stored to replay events. Thus, we focus on the following problem: \textit{Given the amount of permissible overhead for logging events, what are the scenarios where perfect replay of events is possible?} 

\textbf{Contributions of the paper: }We present \newhvc, a reply clock that enables the replay of events in a distributed system. It guarantees that if there is a causal relation \cite{Lamport78CACM} or if $f$ occurred far later than $e$ then $\newhvc.e < \newhvc.f$, i.e., the replay will cause $e$ to replayed before $f$. On the other hand, if $e$ and $f$ are causally concurrent and occurred close in physical time then they could be replayed in any order. By considering various system parameters, clock skew ($\Bigepsilon$), message rate $(\alpha)$, and message delay $(\delta)$, we identify the feasibility region for \newhvc.

\textbf{Organization of the paper: }
The rest of the paper is organized as follows: In Section \ref{sec:Prel}, we describe the model of computation for distributed systems including the notion of causality and clock synchronization. In Section \ref{sec:replay}, we discuss the problem of replay with clocks. Section \ref{sec:replayalgorithm} presents our algorithm for \newhvc.
Properties of \newhvc for solving the replay algorithm are discussed in Section \ref{sec:properties}.
Section \ref{sec:representation} identifies the representation of \newhvc and its overhead. 
We discuss our simulation results in Section \ref{sec:simulation}. Section \ref{sec:related} discusses related work and Section \ref{sec:discussion} identifies questions raised by \newhvc. Finally, in Section \ref{sec:concl}, we conclude and discuss future work. 

\section{Preliminaries}
\label{sec:Prel}
A distributed system is a set of processes $1..n$. Each process has three types of events (1) $send$, where it sends a message to another process, (2) $receive$, where it receives a message from another process, and (3) $local$, where it performs some local computation.



We define the happened-before (denoted by $\hb$) relation \cite{Lamport78CACM} among the events in a distributed computation. 

\begin{itemize}
    \item If $e$ and $f$ happened on the same process and $e$ occurred before $f$  then $e \hb \ f$.
    \item If $e$ was a send event and $f$ was the corresponding receive event  then $e \hb \ f$.
    \item The $\hb$ relation is transitive, i.e., if there exist events $e, f$, and $g$ such that $e \hb \ g$ and $g \hb \ f$ then $e \hb \ f$
\end{itemize}

We say that $e || f$ iff $\neg (e \hb f) \wedge \neg(f \hb\ e)$. In other words, $e$ is concurrent with $f$ iff $e$ did not happen before $f$ and $f$ did not happen before $e$. 


A timestamping algorithm assigns a timestamp for every event $e$ in the system as soon as the event is created. Additionally, the timestamping algorithm defines a $<$ relation that identifies how two timestamps are compared. 

As an illustration, Lamport's logical clock assigns an integer timestamp $l.e$ to every event $e$. The $<$ relation for Lamport's logical clocks is the standard $<$ for integers. 
Likewise, the physical timestamping algorithm assigns $pt.e$ for every event $e$ where $pt.e$ is the physical time of the process where event $e$ occurred when it occurred, and the  $<$ relation is the same as that over integers.
A vector clock \cite{Fidge87}\cite{mattern1988virtual} assigns event $e$ a timestamp $vc.e$ where $vc.e$ is a vector that includes an entry $vc.e.j$ for every process $j$. The $<$ relation on two vector clocks $vc.e$ and $vc.f$ requires that each element in $vc.e$ is less than or equal to the corresponding element in $f$ and some element in $e$ is less than the corresponding element in $f$. In other words, $vc.e < vc.f$ iff $(\forall j :: vc.e.j \leq vc.f.j) \wedge (\exists j :: vc.e.j < vc.f.j)$.

Note that while the $<$ relation is defined by the timestamping algorithm, the properties of the $<$ relation vary. For example, 
logical clocks provide one-way causality information, i.e., $e \hb f \Rightarrow l.e < l.f$. Vector clocks provide two-way causality information, i.e., $e \hb f \Leftrightarrow vc.e < vc.f$. By contrast, (unsynchronized) physical clocks may not provide any guarantees. For example, it is possible that $(e \hb f)$ and $pt.e \not < pt.f$ are simultaneously true.



We assume that each process $j$ in the system is associated with a physical clock, $pt.j$. Clocks of processes are synchronized with a protocol such as NTP \cite{mills1991internet} such that the clock of two processes differ by at most $\Bigepsilon$, where $\Bigepsilon$ is a parameter, i.e., $\forall j, k :: |pt.j -pt.k| \leq \Bigepsilon$. We also assume that individual clocks are monotonically increasing. 
We assume that messages are delivered with a \textit{minimum message delay} of $\delta$. We do not assume maximum message delay; it could be $\infty$ if messages are permitted to be lost. Since our focus is on the replay of events, if a message is lost, it simply implies that the corresponding receive message is never replayed.

\section{Replay with Clocks}
\label{sec:replay}

In this section, we focus on how clocks can be used to replay a given computation. We also discuss some of the limitations of using logical clocks and vector clocks in the replay process. Note that the goal of this section is only to illustrate the concept and the goals of the replay; it does not focus on developing an \textit{efficient} algorithm for the same.  

As discussed in the introduction, the goal of replay is to order the events so that we can evaluate various properties of interest.
To replay a given computation, we begin with the set where each entry is of the form $\br{e, ts.e}$, where $e$ is the event (send/receive/local) and $ts.e$ is the timestamp of $e$. To replay the given set of events, we first find events $e$ such that all events with smaller timestamps than $e$ have already been replayed. In other words, we find the set $\{e | \neg (\exists f: ts.f < ts.e)\}$. We replay one of these events randomly. Then, we remove event $e$. The process is continued until all events are replayed. Thus, the algorithm for replay is shown in algorithm \ref{alg:replay-events}.



\begin{algorithm}
\caption{ReplayEvents Operation}
\label{alg:replay-events}
\begin{algorithmic}[1]
    \State \textbf{Input:} $S$: Set of Events and timestamp
    \While{$S \neq \phi$}
        \State $FrontLine = \{ e | (e, ts.e) \in S \wedge \neg (\exists f: (f, ts.f) \in S \wedge ts.f < ts.e)\}$
        \State Choose a random event $e$ from FrontLine and replay it
        \State $S = S - \{e\}$
    \EndWhile
\end{algorithmic}
\end{algorithm}

\subsection{Limitations of Existing Clocks for Replay}
\label{sec:limitations}
\begin{figure}[ht]
    \centering
    \includegraphics[width=0.6\linewidth]{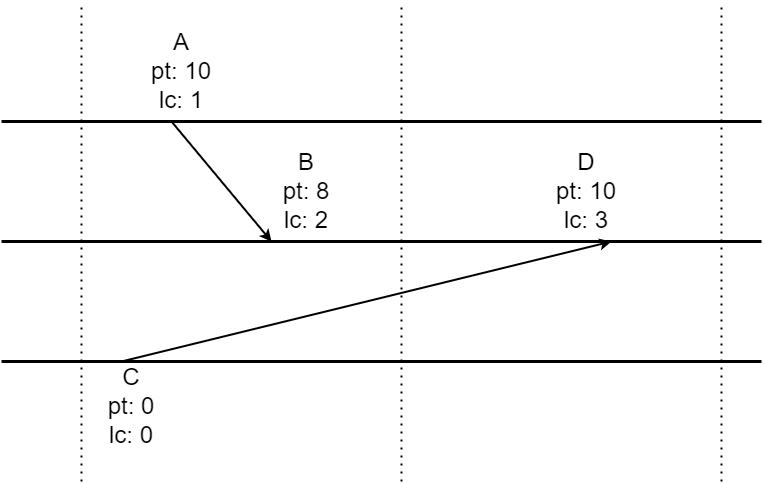}
    \caption{Sample Execution Sequence and Application of Replay Algorithm.}
    \label{fig:intro-example}
\end{figure}

As an example, consider the execution in Figure \ref{fig:intro-example}. Here, we have four events $A, B, C$, and $D$. Their physical timestamps and logical timestamps are shown in Figure \ref{fig:intro-example}. If we replay these events using physical clocks then the possible outcomes are $CBAD$ or $CBDA$. Note that both these outcomes are undesirable, as $A$ should occur before $B$ based on the causality (happened-before) relation. 

If we replay them by logical clocks, the possible outcome is $CABD$. However, there is no option to replay $B$ before $C$ even though $B||C$. 

If we replay them with vector clocks, possible orderings are $ABCD$, $ACBD$, or $CABD$. If the clocks were synchronized to be within 5 time units then $ABCD$ and $ACBD$ are incorrect.

\subsection{Requirements of Replay Clock \newhvc}
\label{sec:newhvcrequirements}

In this paper, we focus on a system where the physical clocks are synchronized to be within $\Bigepsilon$, i.e., for any two processes $j$ and $k$, $|pt.j-pt.k| \leq \Bigepsilon$. The goal of \newhvc is to assign a timestamp $\newhvc.e$ to event $e$ such that 

\begin{description}
    \item [Requirement 1. ] If $e$ happened before $f$ then\\ $\newhvc.e < \newhvc.f$, i.e., $e$ will always be replayed before $f$
    
    \item [Requirement 2. ] If $f$ occurred far after $e$, i.e., $e$ and $f$ could not have occurred simultaneously under clock drift guarantee of $\epsilonone$
    where $\epsilonone 
    \approx 
    \Bigepsilon$ then 
    $e$ will be replayed before $f$, i.e., $\newhvc.e < \newhvc.f$

\item [Requirement 3. ] If $e$ and $f$ could have occurred in any order in a system where clocks were synchronized to be within $\epsilontwo$, where $\epsilontwo\approx\Bigepsilon$ then $\newhvc.e||\newhvc.f$ (i.e., $\neg (\newhvc.e < \newhvc.f) \wedge \neg(\newhvc.f < \newhvc.e))$   

\end{description}

In the last two requirements, we have chosen $\epsilonone$ and $\epsilontwo$ instead of $\Bigepsilon$ itself as it can permit more efficient implementation by allowing us to maintain a coarse-grained clock. We discuss this further in Section \ref{sec:properties}.




\section{Algorithm for Replay Clock (\newhvc)}
\label{sec:replayalgorithm}
In this section, we present our approach for \newhvc. As discussed earlier, we assume that the physical clocks are synchronized to be within $\Bigepsilon$. We discretize the process execution in terms of epochs, where each epoch corresponds to an increment of the clock by $\intervalsize$, $0 < \intervalsize \leq \Bigepsilon$ such that $\Bigepsilon = \epsilon*\intervalsize$, where $\epsilon$ is an integer. 
The timeline of a process is split into epochs where each epoch is of  size \intervalsize (in the local process clock). In other words, the epoch of process $j$ is obtained by $\lfloor \frac{pt.j}{\intervalsize}\rfloor$.


\textbf{Structure of \newhvc timestamp. }
With such discretization, the timestamp of process $j$ (or event $e$) is of the form 
\begin{equation}
    \br{\maxt.j, \texttt{offset}.j[], \texttt{counter}.j[]},
\end{equation}
where $\maxt.j$ is an integer, $\texttt{offset}.j$ and $\texttt{counter}.j$ are arrays that store \textit{at most} one entry $\texttt{offset}.j.k$ and $\texttt{counter}.j.k$ for process $k$



The intuition behind these variables is as follows:
$\maxt.j$ denotes the maximum epoch process $j$ is aware of (either due to the value of $pt.j$ or the value of epochs learned from messages it receives). 
$\maxt.j-\texttt{offset}.j.k$ denotes the maximum epoch value of $k$ that $j$ has learnt (either via direct/indirect message from $k$, clock drift assumption, etc). And, counters are used to deal with the scenario where multiple events happen within the same epoch.

For example, the timestamp $\br{50, [0, 1, 2],[4, 5, 6]}$ denotes that this event is aware of epoch 50 of process $0$, 49 of process $1$, and 48 of process $2$. And, the counter values are $4, 5$ and $6$ respectively. For brevity, we will drop the counter values if they are not relevant to the current discussion. 

\begin{figure}
    \centering
    \includegraphics[width=0.5\linewidth]{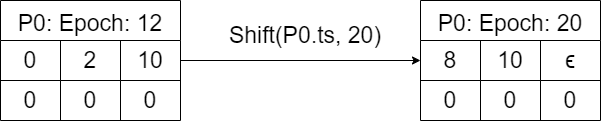}
    \caption{Working of Shift() on Process 0. Here, $\epsilon$=15, and the shift is issued to advance to $\texttt{epoch}$ 20.}
    \label{fig:shift}
\end{figure}

Before we describe the algorithm, we discuss two helper functions, \textit{Shift} and \textit{MergeSameEpoch}. 
The Shift function allows us to change the value of $\maxt$. 
Since $\maxt.j-\texttt{offset}.j.k$ denotes the knowledge of $j$ has about the epoch of $k$, if $\maxt$ is changed to \newmaxt without providing $j$ any additional knowledge of the clock of $k$ then $\newmaxt-\texttt{newoffset}.j.k$ should remain the same as $\maxt.j - \texttt{offset}.j.k$. Hence, Shift operation changes offset$.j.k$ to be $\texttt{offset}.j.k+(\newmaxt-\maxt)$. Furthermore, if this value is more than $\epsilon$ then we reset it to $\epsilon$, as guaranteed by the clock drift assumption. (Note that process $j$ can learn about the clock of $k$ via clock synchronization assumption even if $j$ and $k$ do not communicate. 
For example, shifting the timestamp $\br{12, [0, 2, 10]}$ so that $\maxt$ is changed to 20 will result in $\br{20, [8, 10,18]}$. If $\epsilon=15$, this will change to $\br{20, [8, 10, \epsilon]}$ (cf. Figure \ref{fig:shift}).

The MergeSameEpoch function takes two timestamps $t1$ and $t2$ with the same $\maxt$ value and combines their offsets by setting to be $\texttt{offset}.j.k$ to be the $min(t1.\texttt{offset}.j.k, t2.\texttt{offset}.j.k)$. For example, merging $\br{50, [0, 1, 2]}$ and $\br{50, [2, 0, 1]}$ results in $\br{50, [0, 0,1]}$.

\textbf{Description of the Algorithm. } The algorithm works as follows: 

\textit{Local/Send event. } Here, we describe how $\newhvc.j$ is updated when $j$ sends a message.
Let the current timestamp of $j$ be 
\begin{equation*}
    \br{\maxt.j, \texttt{offset}.j[], \texttt{counter}.j[]}
\end{equation*}
First, $\maxt.j$ needs to be increased if the clock of $j$ has advanced beyond epoch $\maxt.j$. Hence, we first compute $\newmaxt.j$ which is equal to $max(\maxt.j, \texttt{epoch}.j)$. 
When $j$ sends a message, it does not learn any new information about the clock of process $k$. 

We consider two cases: The first case is for the scenario where the newly created event $f$ is in a new epoch as the previous event, $e$. This will happen if $\maxt$ remains unchanged and $\texttt{offset}.j.j$ is unchanged. In this case, we increase $\texttt{counter}.j.j$. 
The second case deals with the scenario where $f$ is in a new interval. Thus, the offset associated with $k$ is changed using the Shift operation. Note that the Shift operation computes the shift of all processes except $j$. $\texttt{offset}.j.j$ should be based on the value of $\texttt{epoch}.j$. Hence, we set it equal to $\newmaxt-\texttt{epoch}.j$. The Shift operation is illustrated in Algorithm \ref{alg:shift}.

\begin{algorithm}
  \caption{Shift Operation}
  \label{alg:shift}
  \begin{algorithmic}[1]
    \State Define Shift operation.
    \State Shift(${ts}, \newmaxt$)
    \For{each $k$}
        \State $ts.\texttt{offset}.k \gets {\texttt{offset}.k} + (\newmaxt - {ts.\maxt})$
        \If{$ts.\texttt{offset}.k > \epsilon$}
            \State $ts.\texttt{offset}.k \gets \epsilon$
        \EndIf
    \EndFor
    
    \State \textbf{Output:} ${ts}$
  \end{algorithmic}
\end{algorithm}



\begin{algorithm}
  \caption{Send Message}
  \begin{algorithmic}[1]
    \State $\newmaxt \gets \max({\maxt.j}, {\texttt{epoch}.j})$
    \State ${\texttt{new\_offset}} \gets {\newmaxt} - {\texttt{epoch}.j}$
    
    \If{$({\maxt.j} = {\newmaxt} \land {\texttt{offset}.j.j} = {\texttt{new\_offset}})$}
      \State ${\texttt{counter}.j.j} \gets {\texttt{counter}.j.j} + 1$
    \Else
      \State ${ts.j} \gets {Shift}({ts.j}, {\newmaxt})$
      \State ${\texttt{offset}.j.j} \gets \min({\newmaxt} - {\texttt{epoch}.j},\epsilon)$
      \State ${\texttt{counter}.j} \gets [0,0,\ldots,0]$
    \EndIf
  \end{algorithmic}
\end{algorithm}

\textit{Receive event. }
Next, we describe how \newhvc is updated when $j$ (with timestamp $\br{\maxt.j, \texttt{offset}.j[], \texttt{counter}.j[]}$ receives a message $m$ with timestamp $\br{\maxt.m, \texttt{offset}.m[], \texttt{counter}.m[]}$.

First, we compute $\newmaxt$ which is the maximum of $\maxt.j$, $\maxt.m$ and $\texttt{epoch}.j$. Timestamps of $j$ and $m$ are then shifted to $\newmaxt$ using the Shift operation. These timestamps are then merged to obtain the $\maxt$ and $\texttt{offset}$ values of the new event, say $f$. 

Now, we check if the knowledge that $f$ has about epochs is the same as that of $e$ (the previous event on $j$) or $m$. If all three are in the same epoch then $\texttt{counter}.j.j$ is set to one more than the maximum of $\texttt{counter}.j.k$ and $\texttt{counter}.m.k$. If only $e$ and $f$ are in the same epoch, $\texttt{counter}.j.j$ is incremented by $1$.
If only $m$ and $f$ are in the same epoch, $\texttt{counter}.j$ is set to $\texttt{counter}.m$ and the value of $\texttt{counter}.j.j$ is incremented by $1$. If none of these conditions apply then counters are reset to $0$. 

\begin{algorithm}
  \caption{Receive Message}
  \begin{algorithmic}[1]
    \State \textbf{Input:} Message $m$
    \State Let $m$ be the message that was received
    \State $\newmaxt \gets \max(\maxt.j, \maxt.m, \texttt{epoch}.j)$
    \State ${ts.a} \gets {\texttt{Shift}}({ts.j}, \newmaxt)$
    \State ${ts.b} \gets {\texttt{Shift}}({ts.m}, \newmaxt)$
    \State ${ts.c} \gets {\texttt{MergeSameEpoch}}({ts.a}, {ts.b})$
    
    \If{EqualOffset$(ts.j, {ts.c}) \land$ EqualOffset$(ts.m, {ts.c})$}
        \For{each $k$}
            \State $\texttt{counter}.j.k \gets \max(\texttt{counter}.j.k, \texttt{counter}.m.k)$
        \EndFor
        \State $\texttt{counter}.j.j \gets \texttt{counter}.j.j + 1$
    \EndIf
    
    \If{$\texttt{EqualOffset}(ts.j, {ts.c}) \land \neg$ $\texttt{EqualOffset}(ts.m, {ts.c})$}
        \State $\texttt{counter}.j.j \gets \texttt{counter}.j.j + 1$
    \EndIf
    
    \If{$\neg$ $\texttt{EqualOffset}(ts.j, {ts.c}) \land$ $\texttt{EqualOffset}(ts.m, {ts.c})$}
        \State $\texttt{counter}.j \gets {\texttt{counter}.m}$
        \State $\texttt{counter}.j.j \gets \texttt{counter}.j.j + 1$
    \EndIf
    
    \If{$\neg$ $\texttt{EqualOffset}(ts.j, {ts.c}) \land \neg$ $\texttt{EqualOffset}(ts.m, {ts.c})$}
        \State $\texttt{counter}.j \gets [0, 0, \ldots, 0]$
    \EndIf
    
  \end{algorithmic}
\end{algorithm}

\begin{figure}[ht]
    \centering
    \includegraphics[width=0.6\linewidth]{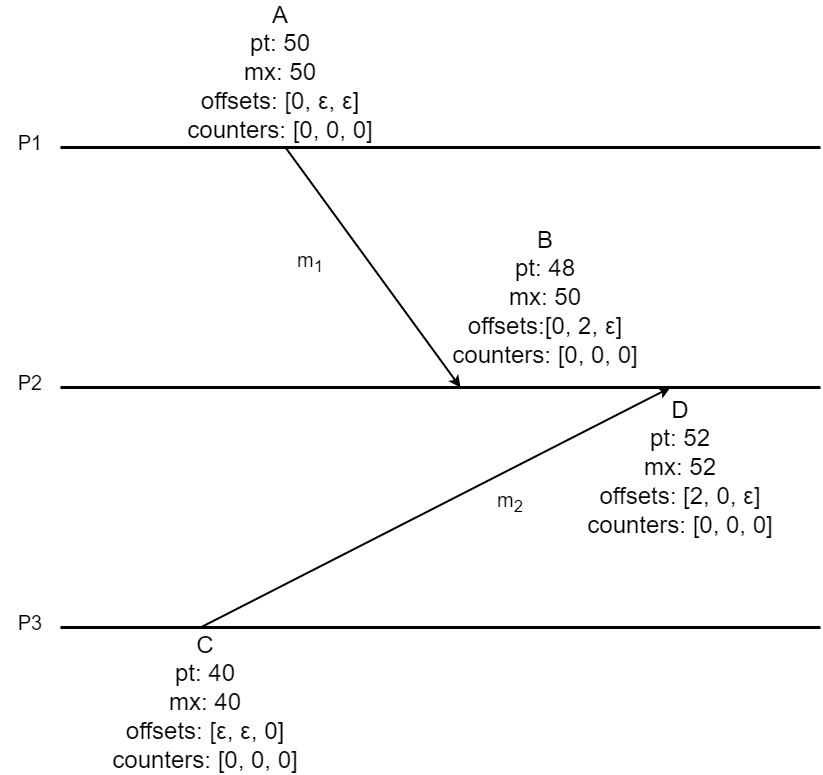}
    \caption{Replay of the Execution in Figure \ref{fig:intro-example} with \newhvc.}
    \label{fig:repcl-example}
\end{figure}

As an illustration, consider the execution of the program in Figure \ref{fig:intro-example}. Assuming that $\epsilon=5,\intervalsize=1,$ and $\Bigepsilon=5$, the \newhvc timestamps will be as shown in Figure \ref{fig:repcl-example}.
Here, event $A$ has physical time of $50$. Since process P1 has not heard from anyone else so far, the offsets for P2 and P3 will be $\epsilon$. The offset for process P1 will be $0$.
Regarding event $C$, the situation is similar except that the offset for process P3 is $0$. 
When event $B$ is created upon receiving message $m_1$, process P2 is aware of times $50$ from $P1$. And, it is the maximum epoch it is aware of. Hence, offsets are $[0, 2,\epsilon]$ respectively. When event $D$ is created, process P2 is aware of epoch 52 (from P2) and epoch 50 (from P1). It is aware of timestamp $40$ from P3. However, this information is overridden by the clock synchronization guarantee that says that the clock of P3 is at least $47$. Thus, the offsets are set to $[3, e, \epsilon]$
Here, the permissible ordering is $CABD$. 

In this figure, if $\epsilon$ were $20$ then the timestamp of $D$ would be changed to $[3,2,12]$. Furthermore, $B$ and $C$ could be replayed in any order. Thus, the permissible replays would be $CABD$ or $ABCD$ or $ACBD$. 


    
\begin{algorithm}
\caption{MergeSameEpoch Operation}
  \begin{algorithmic}[1]
    \State \textbf{Input:} Timestamp $t1$, Timestamp $t2$
    \State Timestamp $ts$ = new Timestamp
    \For{each $k$}
    \State $ts.\texttt{offset}.j.k = min(t1.\texttt{offset}.j.k, t2.\texttt{offset}.j.k)$
    \EndFor
    \State \Return $ts$ 
\end{algorithmic}  
\end{algorithm}


\begin{algorithm}
  \caption{EqualOffset Operation}
  \begin{algorithmic}[1]
  \State \textbf{Input:} Timestamp $t1$, Timestamp $t2$
    \If{$t1.{\maxt} \neq t2.{\maxt} \vee (\exists j \ t1.{\texttt{offset}.j} \neq t2.{\texttt{offset}.j})$}
        \State \Return false
    \Else
        \State \Return true
    \EndIf
  \end{algorithmic}
\end{algorithm}

\section{Properties of \newhvc}
\label{sec:properties}

In this section, first, we define the $<$ relation on two timestamps $\newhvc.e$ and $\newhvc.f$. Then, we identify the properties of this $<$ relation and the happened-before relation. 

\newdefine \newhvcless: \ \ 
    Given timestamp $\newhvc.e = \\
    \br{\maxt.e, \texttt{offset}.e[], \texttt{counter}.e[]}$ and \\
     $\newhvc.f =  \br{\maxt.f, \texttt{offset}.f[], \texttt{counter}.f[]}$, we say that 
     \\$\newhvc.e < \newhvc.f$ iff 

    


\begin{tiny}
\begin{align*}
    \maxt.f &> \maxt.e + \epsilon \\
    \vee &\Big( |\maxt.f - \maxt.e| \leq \epsilon \\
    &\quad \land \Big( \Big( \forall l (\maxt.e - \texttt{offset}.e.l) \leq (\maxt.f - \texttt{offset}.f.l) \Big) \\
    &\quad \land \Big( \exists l (\maxt.e - \texttt{offset}.e.l) < (\maxt.f - \texttt{offset}.f.l) \Big) \Big) \\
    &\quad \vee \Big( \forall l (\maxt.e - \texttt{offset}.e.l) = (\maxt.f - \texttt{offset}.f.l) \\
    &\quad \land \Big( \forall l (\texttt{counter}.e.l) \leq (\maxt.f - \texttt{counter}.f.l) \\
    &\quad \land \exists l (\texttt{counter}.e.l) < (\texttt{counter}.f.l) \Big) \Big) \Big)
\end{align*}
\end{tiny}

The above $<$ relation first compares if $\maxt.f$ and $\maxt.e$ are far apart. If that is the case, we define $\newhvc.e < \newhvc.f$. If they are close, i.e., $|\maxt.f-\maxt.e|\leq \epsilon$, then, we compare the offsets. Since $\maxt.e-$offset$.e.k$ identifies the knowledge $e$ had about the epoch of process $k$, we use a comparison that is similar to vector clocks to determine if $<$ relation holds between $\newhvc.e$ and $\newhvc.f$. Finally, if the offsets are also equal then we use the comparison of counters (again in the same fashion as vector clocks).

We overload the $||$ relation for comparing timestamps as well.  Specifically,

\newdefine \concurtimestamps\ \  Given timestamps $\newhvc.e$ and $\newhvc.f$, we say that $\newhvc.e || \newhvc.f$ iff $\neg(\newhvc.e < \newhvc.f) \wedge \neg(\newhvc.f < \newhvc.e)$    

From the construction of the timestamp algorithm, we have the following two lemmas:


\newlem \newhvchappenedbefore: \ \ 
    (e happened before f) 
    $\Rightarrow \newhvc.e < \newhvc.f$

\newlem \newhvcconcur: \ \ 
$|\maxt.e-maxt.f|\leq \epsilon \wedge (e||f)\  \ \ \Rightarrow \ \ \  \newhvc.e || \newhvc.f$

\noindent \textbf{Requirement 1 of \newhvc: }Observe that Lemma \newhvchappenedbefore satisfies the first requirement of $\newhvc$; if $e$ happened before $f$ then $e$ must be replayed before $f$. 

\noindent \textbf{Requirement 2 of \newhvc: }
Now, we focus on the second requirement. Specifically, we show that by letting $\epsilonone=\Bigepsilon+\intervalsize$, the second requirement is satisfied.

Observe that in the $\newhvc$ algorithm, messages carry the epoch values of multiple processes. This allows a process to learn epoch information about other processes. For the subsequent discussion, imagine that the messages also carried the actual physical time as well. In this case, $j$ will learn about the clock of a process $k$ via such messages. Additionally, $j$ will also learn about the clock of a process $k$ based on the assumption of clock synchronization. Likewise, when event $e$ is created, it will have some information about the clock of each process. 
Let $\maxph.e$ and $\maxph.f$ be the maximum clock (of any process) that $e$ and $f$ are aware of when they occurred. 
If $\maxph.f > \maxph.e+\epsilonone$ then $f$ cannot occur before $e$ under the clock synchronization guarantee of $\epsilonone$. Now, we show that in this situation, it is guaranteed that $\newhvc.e < \newhvc.f$.

By definition of $\maxt$, $\maxt.f = \lfloor \frac{\maxph.f}{\intervalsize} \rfloor$ and  $\maxt.e = \lfloor \frac{mpt.e}{\intervalsize} \rfloor$. Additionally, we have 

\begin{tabbing}
    \hspace*{5mm} \=
    $-1 < (x - \lfloor x \rfloor) - (y - \lfloor y \rfloor) < 1$\\
$\therefore$  \> 
    $-1 < (\frac{\maxph.f}{\intervalsize} - \lfloor \frac{\maxph.f}{\intervalsize}  \rfloor) - (\frac{\maxph.e}{\intervalsize} - \lfloor \frac{\maxph.e}{\intervalsize} \rfloor) < 1$\\
$\therefore$\>
    $-1 < (\frac{\maxph.f}{\intervalsize} - \maxt.f) - (\frac{\maxph.e}{\intervalsize} - \lfloor \frac{\maxph.e}{\intervalsize} \rfloor) < 1$\\
$\therefore$ \> 
    $-1 < (\frac{\maxph.f-\maxph.e}{\intervalsize}) - (\maxt.f - \maxt.e)  < 1$\\
\end{tabbing}

Now, if $\maxph.f-\maxph.e > \Bigepsilon+\intervalsize$ then we can rewrite the second inequality as 
    $(\frac{\Bigepsilon+ \intervalsize}{\intervalsize} - (\maxt.f - \maxt.e)  \leq 1$. Using the fact that $\Bigepsilon=\epsilon*\intervalsize$, we have $\epsilon < (\maxt.f-\maxt.e)$. In other words, $\maxph.f-\maxph.e > \Bigepsilon + \intervalsize \Rightarrow (\maxt.f > \maxt.e+\epsilon)$. By Definition \newhvcless, we have $\newhvc.e < \newhvc.f$. In other words, we have


\newlem \effar: \ \ 
\label{lem3}
    If $f$ occurred far after $e$, i.e., $f$ could not have occurred before $e$ in a system that guarantees that clocks are synchronized within $\epsilonone = \Bigepsilon+\intervalsize$ then $e$ will be replayed before $f$, i.e., $\newhvc.e < \newhvc.f$. 

\noindent
\textbf{Requirement 3 of \newhvc. }
Next, we consider the case where $e$ and $f$ could have occurred in any order if the underlying system guaranteed that clocks were synchronized to be within $\epsilontwo=\Bigepsilon-\intervalsize$. Letting the maximum clock that event $e$ (respectively, $f$) was aware of to be $\maxph.e$ (respectively, $\maxph.f$), we observe that $|\maxph.e - \maxph.f| \leq \epsilontwo$. Furthermore, $e$ and $f$ must be causally concurrent. Under this scenario, we show that $\newhvc.e || \newhvc.f$. 

If $|\maxph.e - \maxph.f| \leq \Bigepsilon-\intervalsize$, we have

\begin{tabbing}
    \hspace*{5mm} \=

$|mpt.e - mpt.f| \leq \Bigepsilon - \intervalsize$\kill
$\therefore$ \>
${|\frac{mpt.e}{\intervalsize} - \frac{mpt.f}{\intervalsize}| \leq \ \epsilon-1}$ \hspace*{5mm} \= // since $\Bigepsilon=\epsilon*\intervalsize$ \\
$\therefore$\>
${|\lfloor \frac{mpt.e}{\intervalsize} \rfloor - \lfloor \frac{mpt.f}{\intervalsize} \rfloor| \leq \ \epsilon}$ \>
since $|(x - \lfloor x \rfloor) - (y - \lfloor y \rfloor)| < 1$\\
$\therefore$
\> $|\maxt.e-\maxt.f|\leq \epsilon$
\> by definition of $\maxt$
\end{tabbing}

Now, from Lemma 2, $\newhvc.e || \newhvc.f$. In other words, 


\newlem \allorder
If $e$ and $f$ could have occurred in any order in a system where clocks were synchronized to be within $\Bigepsilon-\intervalsize$ then $\newhvc.e||\newhvc.f$.    

\vspace*{2mm}
\noindent \textbf{Effect of discretization and comparison with hybrid vector clocks \cite{yingchareonthawornchai2018analysis}}:
We note that the discretization of the clock via $\intervalsize$ has caused the bounds used for clock synchronization in Lemmas 1 and 2 to be different. We could have eliminated this if we had not discretized the clocks. (Discretization with $\intervalsize$ was not done in \cite{yingchareonthawornchai2018analysis}.)
However, without discretization, the values of offsets will be very large. Without discretization, we will need to rely on just the physical clocks which have a granularity of under 1 nanosecond. Now, if $\Bigepsilon=1ms$ then the value of the offset could be as large as $10^6$. By discretizing the clock, it would be possible to keep offsets to be very small (about 4 bits). We expect that the discretization will not seriously impact the replay. For example, if $\Bigepsilon=1ms$ and $\intervalsize=0.1ms$ then our algorithm will guarantee that causally concurrent events within $0.9ms$ can be replayed in any order. And, events that could not occur simultaneously under a clock synchronization guarantee of $1.1ms$ will be replayed only in one order. Additionally, if $e$ happened before $f$ then $e$ will always be replayed before $f$. 

\section{Representation of \newhvc and its Overhead}
\label{sec:representation}


In this section, we identify how $\newhvc$ can be stored for permitting efficient computation. As written, $\newhvc$ will require $2n+1$ integers. However, a more compact representation will be possible when we account for the fact that it is being used in a system where the clocks are synchronized to be within $\Bigepsilon$. Thus, if $j$ does not hear from $k$ (directly or indirectly) for a long time then the knowledge $j$ would have about the clock of $k$ is the same that is provided by the clock synchronization assumption. In this case, offset$.j.k=\epsilon$. It follows that there is no need to store this value if we interpret \textit{no information about the offset of $k$} to mean that offset$.j.k=\epsilon$.

\begin{figure}
    \centering
    \includegraphics[width=0.9\linewidth]{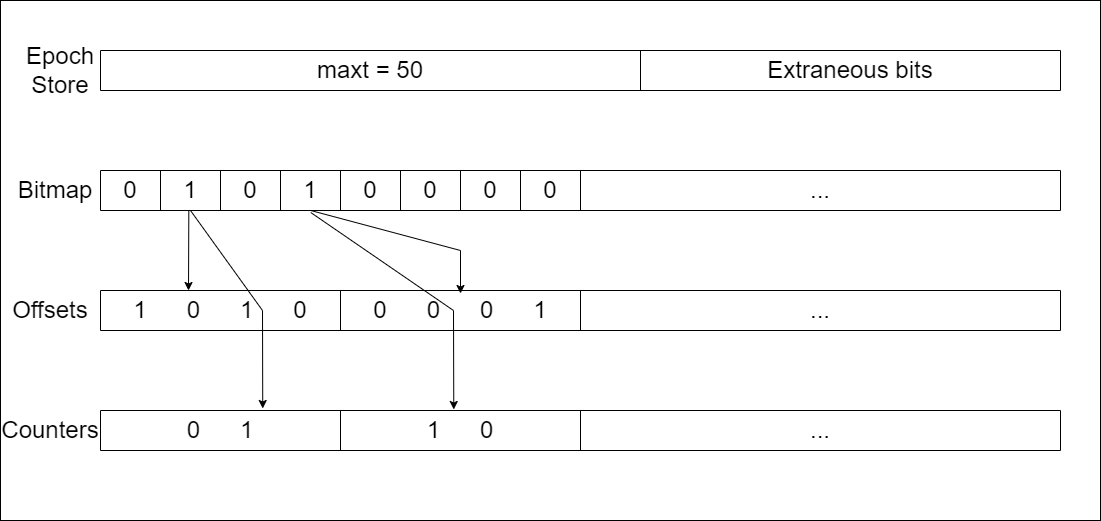}
    \caption{Sample $\newhvc$ representation.}
    \label{fig:size_repr}
\end{figure}

With this intuition, we represent $\newhvc.j$ as shown in Figure \ref{fig:size_repr}. Here, the value of $\maxt.j$ (represented by the first word) is 50. The next word identifies the bitmap. Since the bit corresponding to process 1 is $0$, it implies that offset$.j.0=\epsilon$. The offset for process 2 is $10$ (first 4 bits of the offset) and $counter.j.2$ is $2$ (first 2 bits of the counter). We note that the bits for each offset and counter is hard-coded based on the system parameters (cf. Section \ref{sec:simulation}).
The second word is a bitmap that identifies whether offset$.j.k$ is stored for process $k$. Each offset is stored with a fixed number of bits in the subsequent word(s). 

Next, we show that this representation allows us to reduce the cost of storage as well as the cost of computing timestamps or comparing them (using $<$ relation). Specifically, all these costs are proportional to the number of processes that have communicated with $j$ recently. 
 
With representation in Figure \ref{fig:size_repr}, first, we note that finding the location of the $1$s in the given bitmap can be done in time that is proportional to the number of $1$s in the bitmap. ($(n - (n \& (n - 1)$ will return the number with only the rightmost $1$). Thus, we have

\observe \ShiftEpochComplexity: \ \ 
Shift and MergeSameEpoch can be implemented using $O(x)$ time where $x$ is the number of bits set to $1$ in the bitmap.

Note that this means that the time to compute the timestamp for send/receive at process $j$ is not dependent upon the number of processes in the system. But only processes that have recently communicated with process $j$. In turn, this means that

\observe \SendReceiveComplexity: \ \ 
Send and Receive can be implemented using $O(x)$ time where $x$ is the number of bits set to $1$ in the bitmap.

\observe \Comparetimestamps: \ \ 
Given two timestamps, $\newhvc.e$ and $\newhvc.f$, we can determine if $\newhvc.e < \newhvc.f$ in $O(x)$ time where $x$ is the number of bits set to $1$ in $e$ and $f$. 

It follows that the number of bits that are $1$ in a given timestamp identifies not only the storage cost of the timestamps but also the time to compute these timestamps at run time. Effectively, this also identifies the overhead of the timestamps that enable the replay of the computation. Hence, in Section \ref{sec:simulation}, we focus on identifying scenarios where the cost of storing these offsets is within the limits identified by the user. 

We note that the above approach will work as long as the number of processes is less than the number of bits in a word (typically, 64 in today's systems). We expect that this will be more than sufficient for many systems in practice. If there are more than 64 processes, we expect that process $j$ is communicating with only a subset of these processes. And, if process $j$ does not communicate with someone there is no need to store offsets for them. Thus, this approach can be extended for the case where the number of processes is larger. However, the details are out of the scope of the paper. 

\section{Simulation Results}
\label{sec:simulation}

As demonstrated in Section \ref{sec:properties}, the overhead of $\newhvc$ depends upon the number of offsets/counters that need to be stored. And, this value depends upon the number of processes that communicate with a given process in the $\Bigepsilon$ time. In other words, the system parameters will determine the size of $\newhvc$. In this section, we evaluate the overhead of $\newhvc$ via simulation.


\begin{figure*}[ht]
  \centering

  \begin{subfigure}{0.3\textwidth}
    \includegraphics[width=\linewidth]{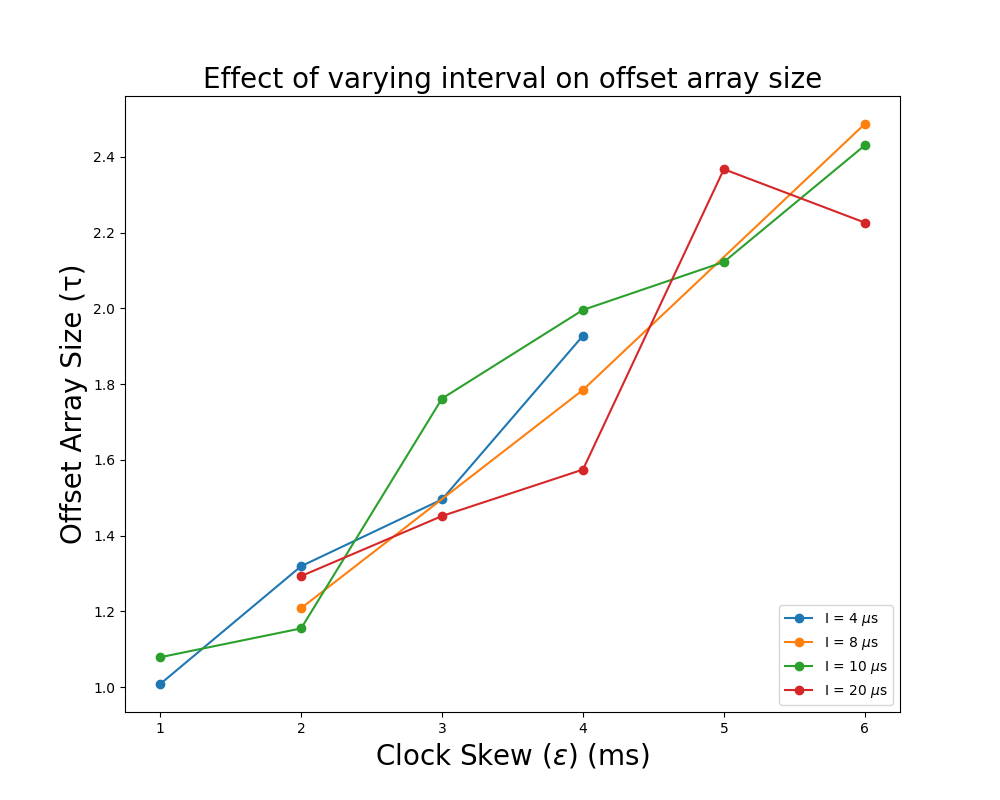}
    \caption{$\alpha$ = 20 messages/s, $n$ = 32.}
    \label{fig:effectofeps-interval1}
  \end{subfigure}
  \hfill
  \begin{subfigure}{0.3\textwidth}
    \includegraphics[width=\linewidth]{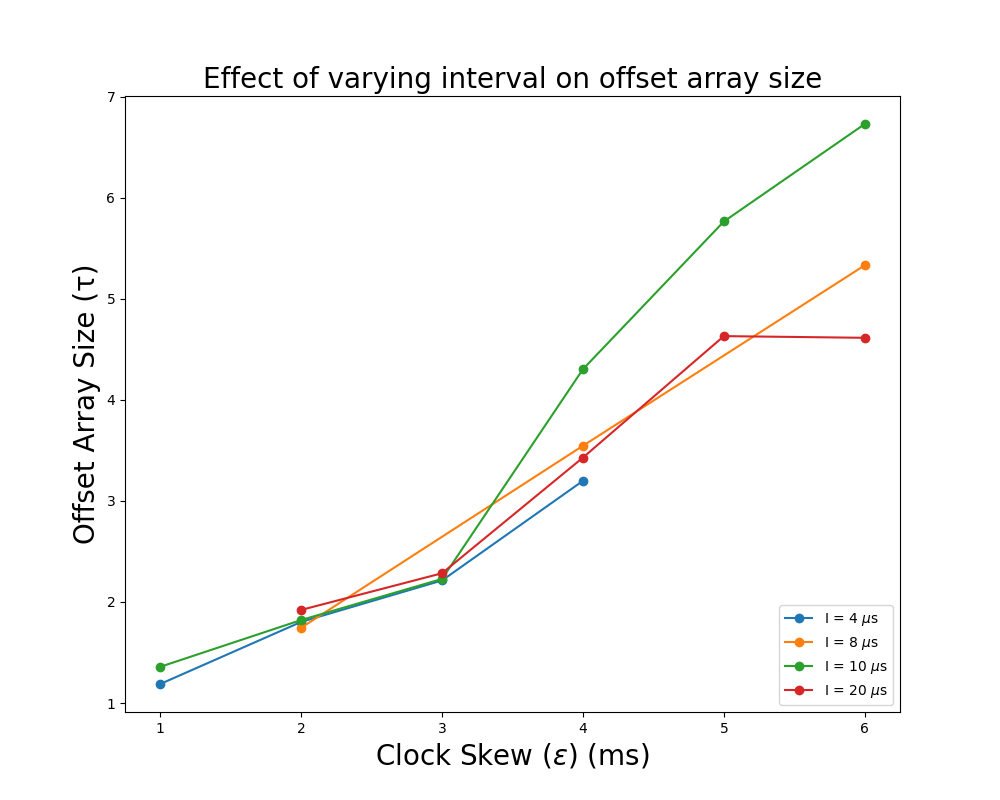}
    \caption{$\alpha$ = 40 messages/s, $n$ = 32.}
    \label{fig:effectofeps-interval2}
  \end{subfigure}
  \hfill
  \begin{subfigure}{0.3\textwidth}
    \includegraphics[width=\linewidth]{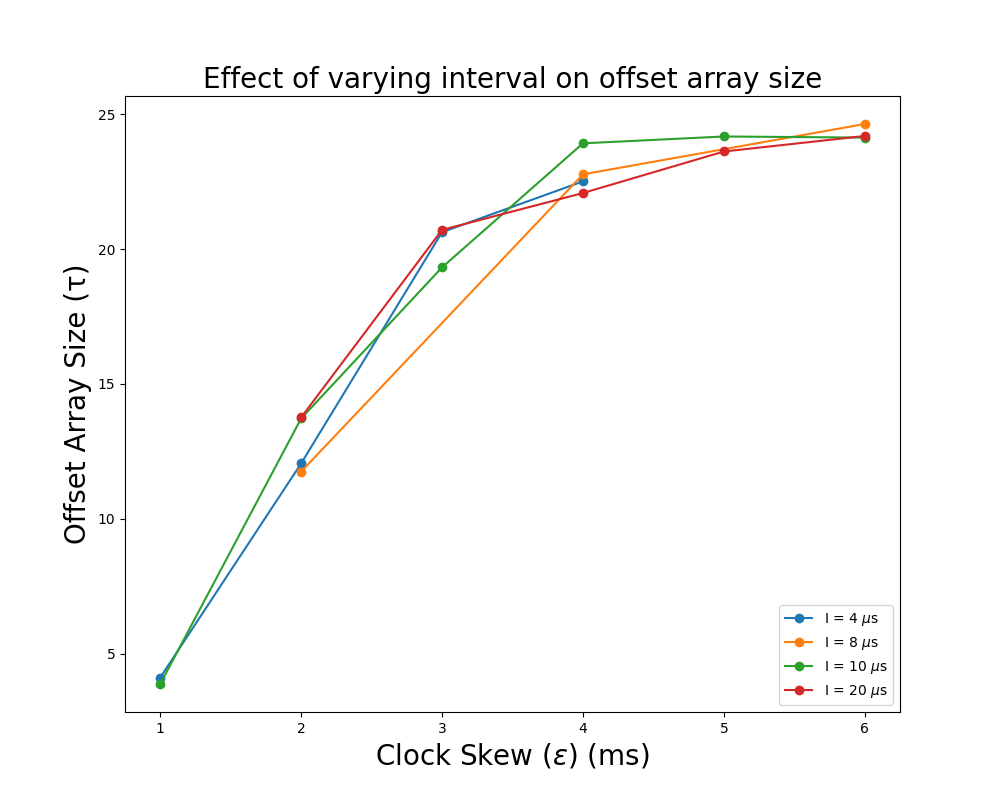}
    \caption{$\alpha$ = 160 messages/s, $n$ = 32.}
    \label{fig:effectofeps-interval3}
  \end{subfigure}

  \begin{subfigure}{0.3\textwidth}
    \includegraphics[width=\linewidth]{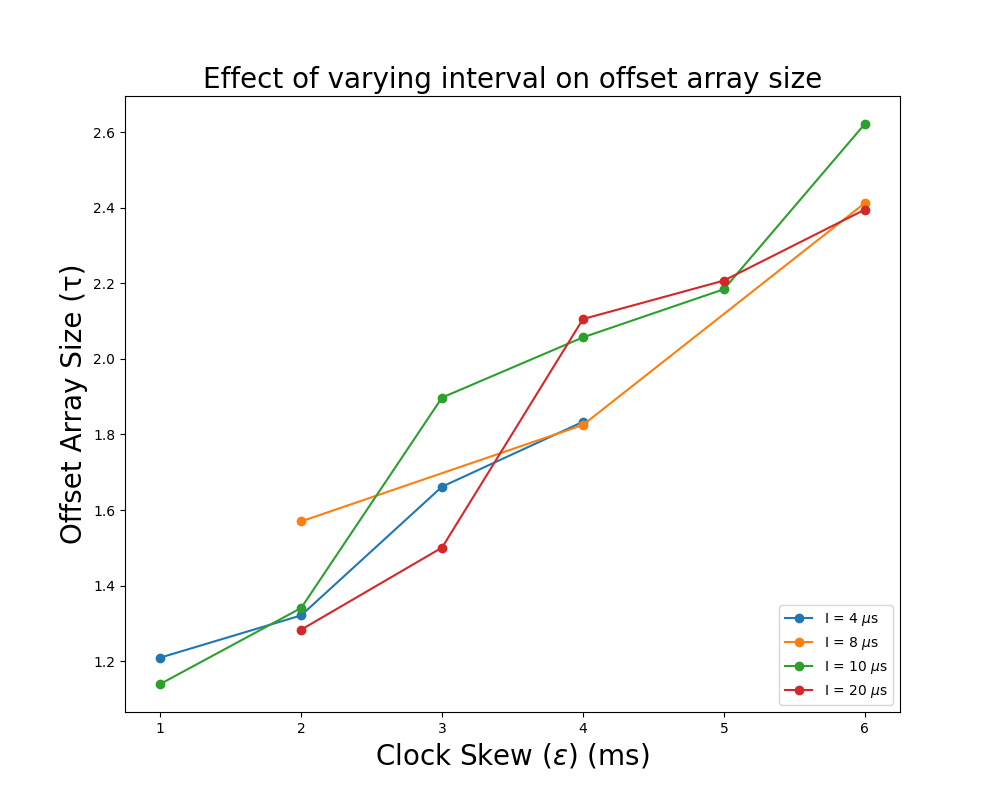}
    \caption{$\alpha$ = 20 messages/s, $n$ = 64.}
    \label{fig:effectofeps-interval4}
  \end{subfigure}
  \hfill
  \begin{subfigure}{0.3\textwidth}
    \includegraphics[width=\linewidth]{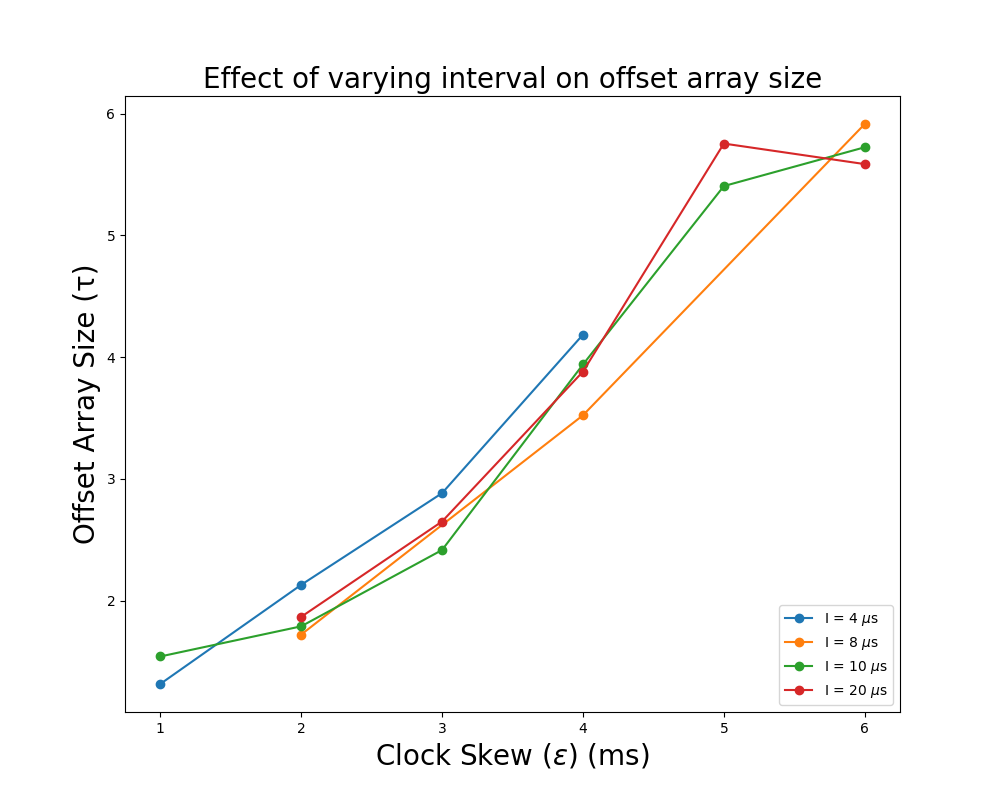}
    \caption{$\alpha$ = 40 messages/s, $n$= 64.}
    \label{fig:effectofeps-interval5}
  \end{subfigure}
  \hfill
  \begin{subfigure}{0.3\textwidth}
    \includegraphics[width=\linewidth]{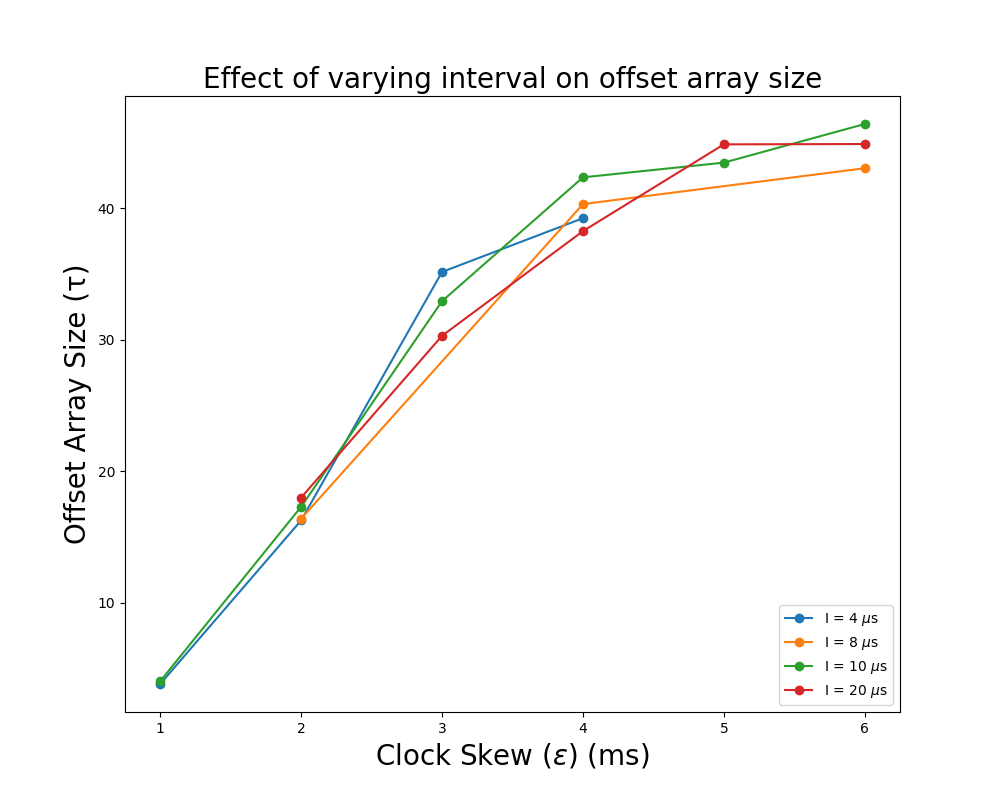}
    \caption{$\alpha$ = 160 messages/s, $n$ = 64.}
    \label{fig:effectofeps-interval6}
  \end{subfigure}

  \caption{$\offsetsize$ vs $\Bigepsilon$ when varying $\intervalsize$, $\delta = 8 \mu s$.}
  \label{fig:effectofeps-intervaloffset}
\end{figure*}

Specifically, we simulate a distributed environment with five key parameters - the number of processes ($n$), the maximum allowed clock skew ($\Bigepsilon$), the interval size ($\intervalsize$), the message rate ($\alpha$), and the message delay in microseconds ($\delta$). The message delay is modeled as the average delay experienced in the simulated network, and can vary by some nonzero $\Delta$ in production.
In the simulation, at every \textit{clock tick}, a process delivers any messages it is expected to deliver at that clock tick. It also sends a message to other processes based on the message rate $\alpha$. When a message is sent, the corresponding receive event is added to the receiver's queue based on the value of $\delta$. Clock skew of $\Bigepsilon$ is enforced by ensuring that a process advances its clock only if it will keep the clocks to be within $\Bigepsilon$ interval. We also compute the actual value of the maximum clock skew observed in the simulation to ensure that if $\Bigepsilon=1ms$ then the worst-case clock skew is indeed $1ms$. 
For the purposes of the results, we denote the offset array size as $\offsetsize$ and the counter array size as $\countersize$.

Next, we describe our observations from these simulations. \\Specifically, we identify how $\offsetsize$ and $\countersize$ change based on varying $\alpha, \delta, \Bigepsilon$ and $\intervalsize$.

\subsection{Effect of Clock Skew (\texorpdfstring{$\Bigepsilon$}{epsilon})}

In this section, we measure the trends in $\Bigepsilon$ while varying the other parameters to see how the $\offsetsize$ and $\countersize$ are affected. We compare each parameter pair-wise with $\Bigepsilon$ to see the effect of the parameter on the clock skew trend with $\offsetsize$.

\begin{figure}[ht]
  \centering

  \begin{subfigure}{0.22\textwidth}
    \includegraphics[width=\linewidth]{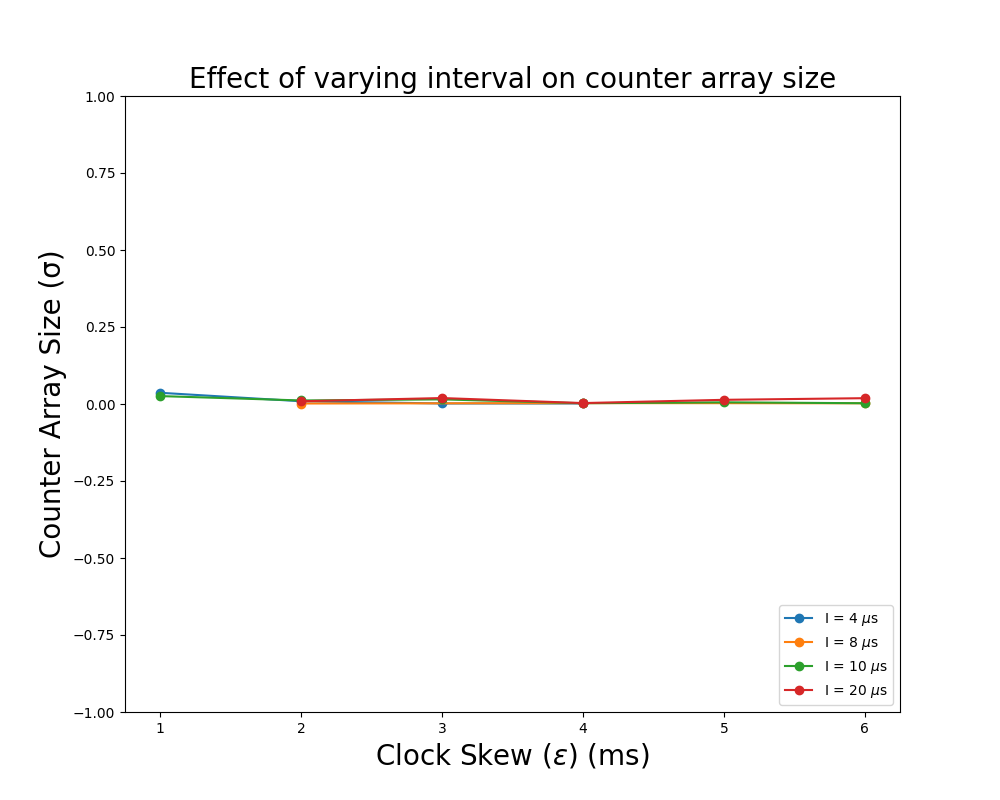}
    \caption{$n$ = 32.}
    \label{fig:effectofeps-interval7}
  \end{subfigure}
  \begin{subfigure}{0.22\textwidth}
    \includegraphics[width=\linewidth]{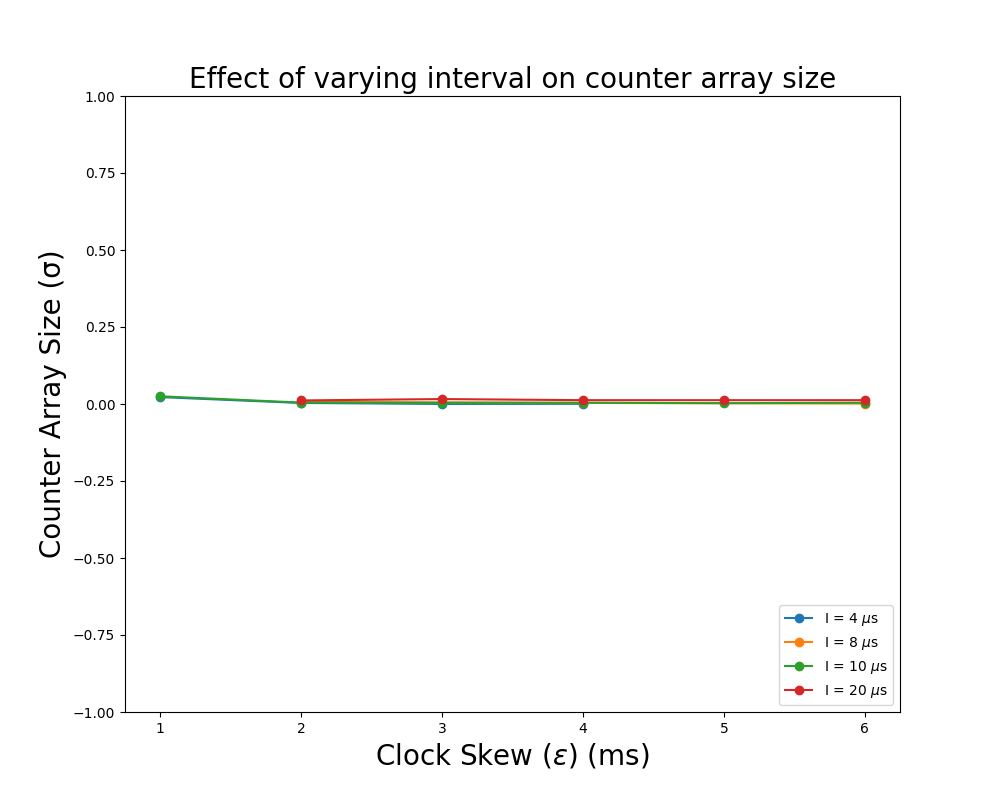}
    \caption{$n$ = 64.}
    \label{fig:effectofeps-interval8}
  \end{subfigure}

  \caption{$\countersize$ vs $\Bigepsilon$ when varying $\intervalsize$, $\delta$ = 8 $\mu s$, $\alpha$ = 160 messages/s.}
  \label{fig:effectofeps-intervalcounter}
\end{figure}

\textbf{Varying $\intervalsize$: } Here, we keep the $\delta$ and $\alpha$ constant to see how $\offsetsize$ and $\countersize$ change with $\Bigepsilon$. (1) In case of $\offsetsize$ vs $\Bigepsilon$ curve, we notice that the value of $\intervalsize$ has little bearing on $\offsetsize$.
As expected, as the value of $\Bigepsilon$ increases, $\offsetsize$ increases with it. This is true for all values of $\alpha$, and we consistently store more offsets as $\alpha$ increases. For any given value of $\alpha$, however, the value of $\intervalsize$ can be chosen to set the granularity of the user's choice and would allow more flexibility in the clock information. Regardless of the choice of $\intervalsize$ by the user, the offset sizes increase with roughly the same trend. These trends are illustrated in Figure \ref{fig:effectofeps-intervaloffset}. 
(2) In the case of $\sigma$ vs $\Bigepsilon$ curve, we see not too much of a variation in $\countersize$, as most events reach a different epoch. 
On average, we do not see many events storing counters; roughly $0.78\%$ of events store counters, and the values of such counters do not exceed 5 in most cases. 
Hence, we would need very little space to store these counters. These trends are illustrated in Figure \ref{fig:effectofeps-intervalcounter}. Since this observation is true for all simulations in this paper, we do not discuss the analysis for $\countersize$ in the subsequent sections.

Since the total size of $\newhvc$ depends upon the number of bits for each offset and the total number of offsets, we consider a specific example here. For Figure \ref{fig:size_repr}, the number of offsets is 2 and the size of each offset is 4 bits. Therefore, one word is sufficient to store offsets. Likewise, one word is enough for counters. Thus, we need a total of 4 words to store this timestamp.
(Note that the counters can be stored in the same amount of memory as we have a number of extraneous bits in this representation, specifically in the max epoch word if we elect to store the sum of all these counters here. By doing this, we would lose some information, but considering that the number of events that record meaningful counters is low, this may be an acceptable trade-off.) It is straightforward to observe that the size of $\newhvc$ grows linearly with the number of offsets in it. And, the total size of $\newhvc$ will require the use of the floor function to identify the number of words necessary to store it. Since the floor operation loses some of the relevant data, we present the value of $\tau$ in this section. 
Graphs that compute the total size of $\newhvc$ are presented in the Appendix.

\begin{figure}[ht]
  \centering

  \begin{subfigure}{0.22\textwidth}
    \includegraphics[width=\linewidth]{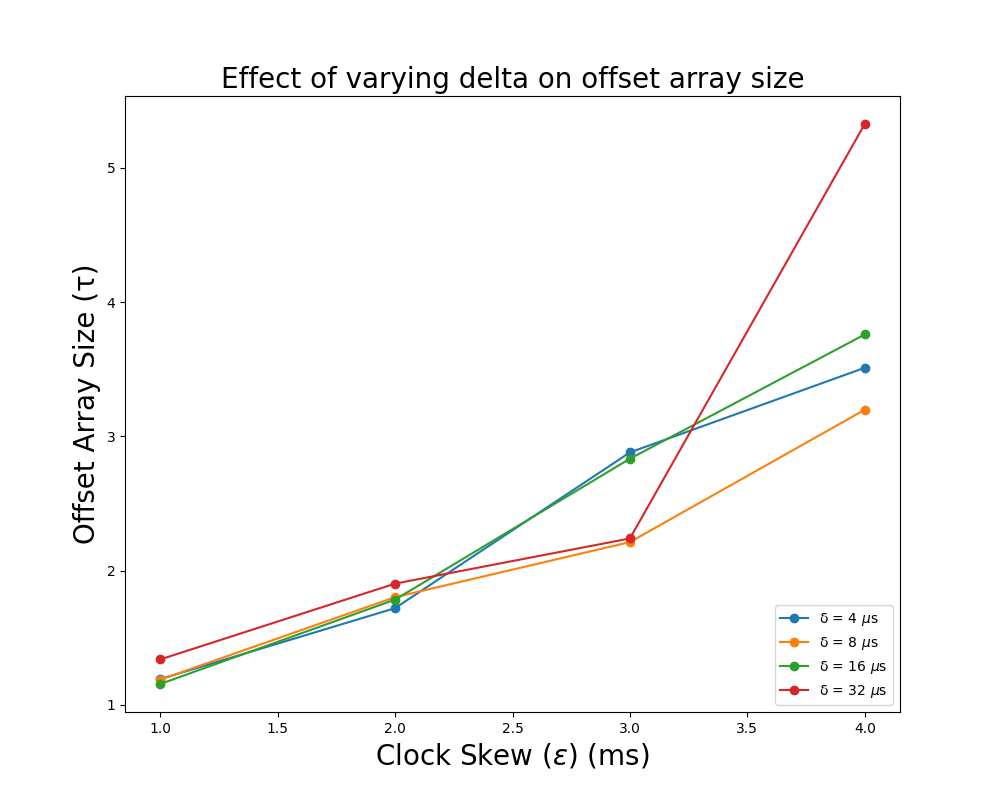}
    \caption{$n$ = 32.}
    \label{fig:effectofeps-delta1}
  \end{subfigure}
  \hfill
  \begin{subfigure}{0.22\textwidth}
    \includegraphics[width=\linewidth]{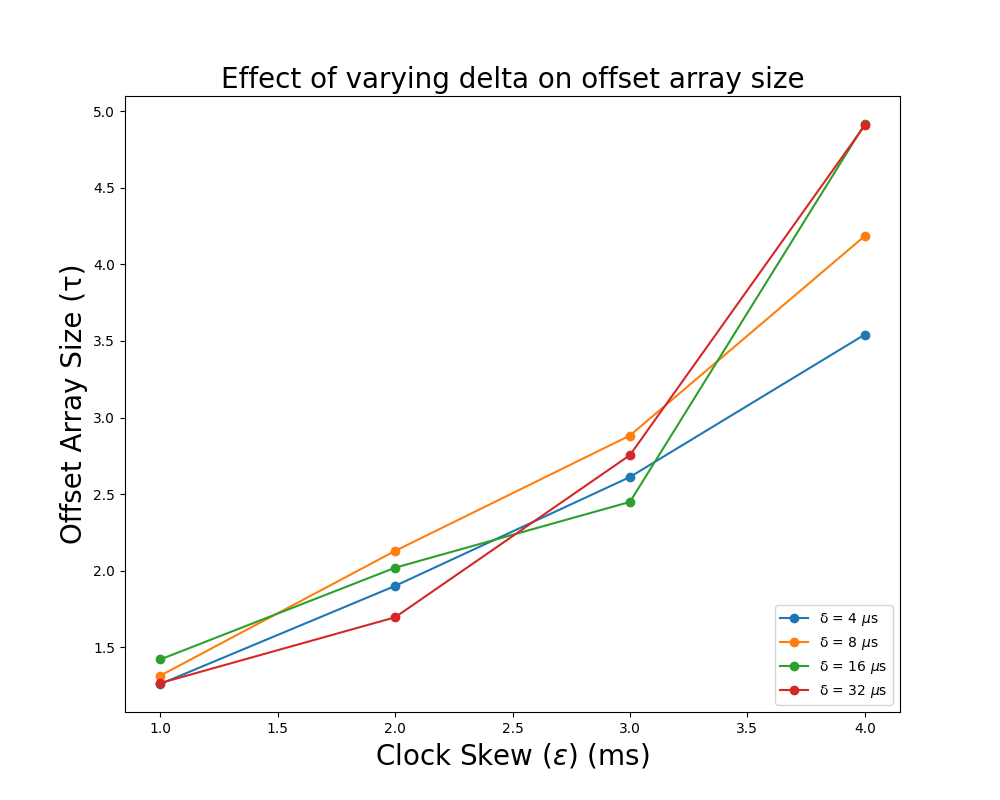}
    \caption{$n$ = 64.}
    \label{fig:effectofeps-delta2}
  \end{subfigure}

  \caption{$\offsetsize$ vs $\Bigepsilon$ when varying $\delta$, $\intervalsize$ = 8 $\mu s$, $\alpha$ = 40 messages/second.}
  \label{fig:effectofeps-deltaoff}
\end{figure}

\textbf{Varying $\delta$: } Here,  we fix the $\intervalsize$ and $\alpha$, and for different $\delta$ values, and we identify how $\offsetsize$ changes with $\Bigepsilon$.
As $\Bigepsilon$ increases, we see higher values of $\offsetsize$, implying a higher number of offsets stored on average. We observe that higher values of $\delta$ produce a lower number of offsets in each case, barring some noise. This is expected as an increase in $\delta$ implies messages would reach in a delayed fashion, and would lead to processes setting other process offsets to $\epsilon$ due to non-receipt of messages.
As the $\Bigepsilon$ increases, a process hears from more processes (directly or indirectly) within time $\Bigepsilon$. Hence, the number of offsets increases.
This is illustrated in Figure \ref{fig:effectofeps-deltaoff}. 

\begin{figure}[ht]
  \centering

  \begin{subfigure}{0.22\textwidth}
    \includegraphics[width=\linewidth]{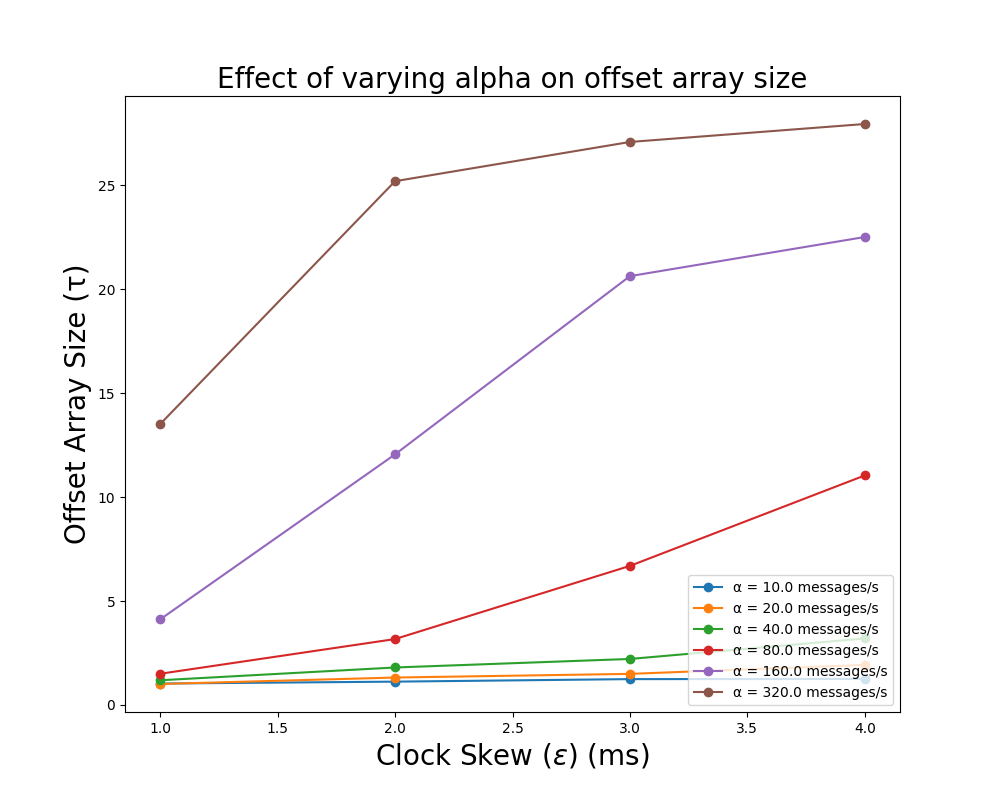}
    \caption{$n$ = 32.}
    \label{fig:effectofeps-alpha1}
  \end{subfigure}
  \hfill
  \begin{subfigure}{0.22\textwidth}
    \includegraphics[width=\linewidth]{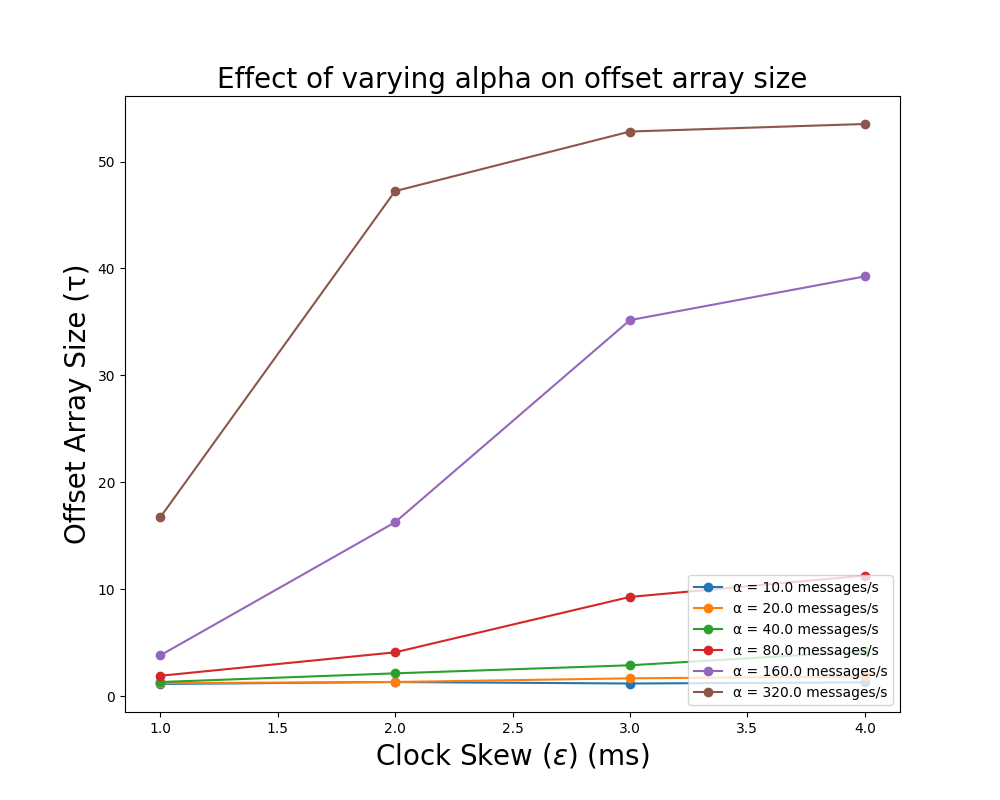}
    \caption{$n$ = 64.}
    \label{fig:effectofeps-alpha2}
  \end{subfigure}

  \caption{$\offsetsize$ vs $\Bigepsilon$ when varying $\alpha$, $\intervalsize$ = 4 $\mu s$, $\delta$ = 8 $\mu s$.}
  \label{fig:effectofeps-alphaoff}
\end{figure}

\textbf{Varying $\alpha$: }Here, we fix the $\intervalsize$ and $\delta$, and for different $\alpha$ values, we identify how $\tau$  changes with $\Bigepsilon$.
As expected, for lower values of $\alpha$, we consistently store lower offsets, as communication between processes is sporadic. As the $\Bigepsilon$ increases, $\offsetsize$ increases linearly until the bound of $n$ is reached. This is due to the same reason mentioned earlier, as the bound lengths on epochs are larger, even sporadic messages tend to store more offsets on other processes, causing the overall value of $\offsetsize$ to increase. This is illustrated in Figure \ref{fig:effectofeps-alphaoff}.

\subsection{Effect of Interval Size (\texorpdfstring{$\intervalsize$}{I})}

In this section, we observe the trends in $\intervalsize$ with respect to $\delta$ and $\alpha$. 

\begin{figure}[ht]
  \centering

  \begin{subfigure}{0.22\textwidth}
    \includegraphics[width=\linewidth]{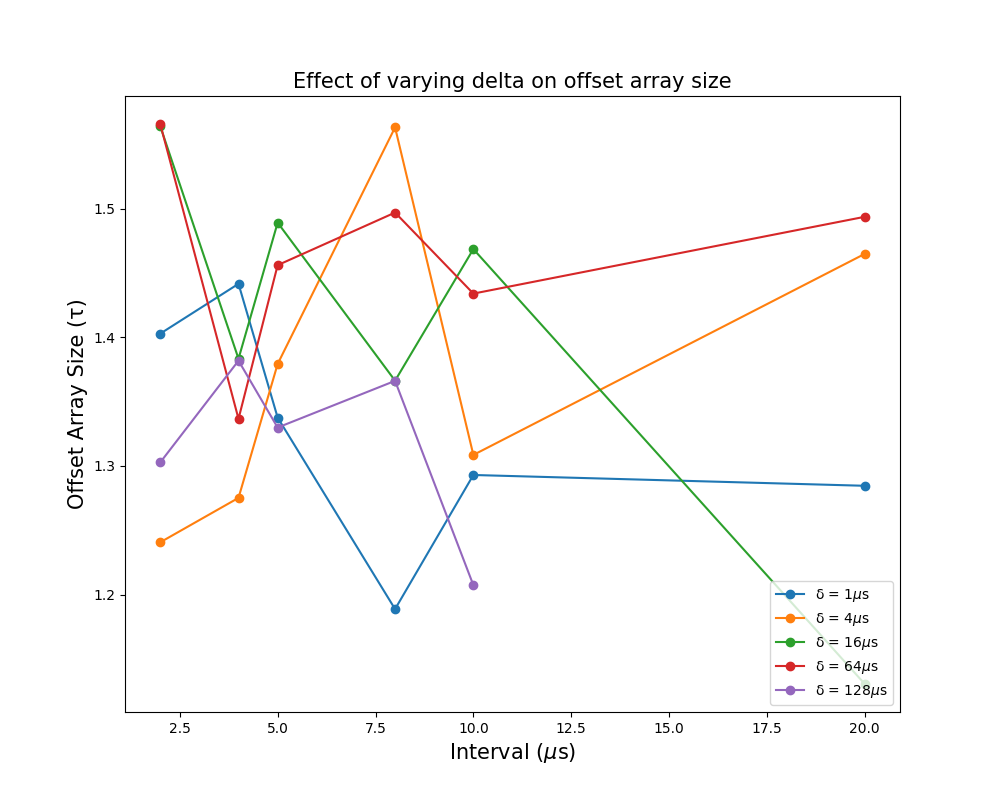}
    \caption{$n$ = 32.}
    \label{fig:effectofint-delta1}
  \end{subfigure}
  \hfill
  \begin{subfigure}{0.22\textwidth}
    \includegraphics[width=\linewidth]{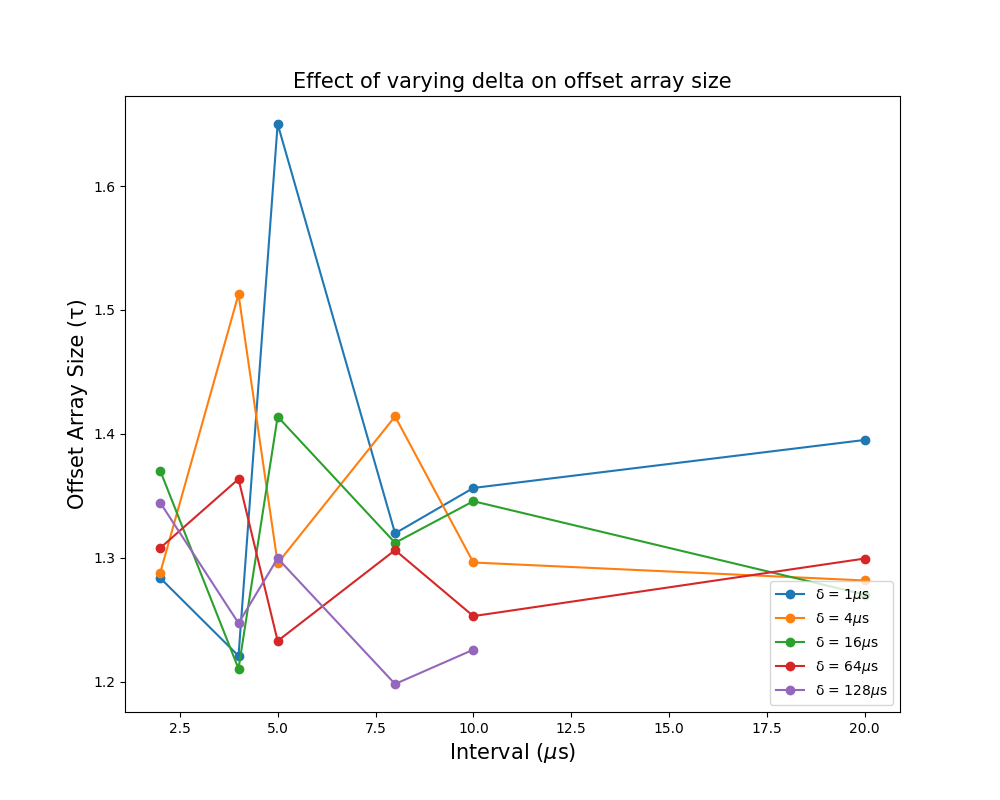}
    \caption{$n$ = 64.}
    \label{fig:effectofint-delta2}
  \end{subfigure}

  \caption{$\offsetsize$ vs $\intervalsize$ when varying $\delta$, $\Bigepsilon$ = 2 ms, $\alpha$ = 20 messages/s.}
  \label{fig:effectofint-deltaoff}
\end{figure}

\textbf{Varying $\delta$: }
Here, we fix the $\Bigepsilon$ and $\alpha$ and check how $\tau$ changes with $\intervalsize$. 
From Figure \ref{fig:effectofint-deltaoff}, we observe that the value of $\intervalsize$ does not really have a significant effect on $\tau$ (Note that the $Y$ axis of this figure varies only from $1.2$ to $1.5$.) This means that the selection of $\intervalsize$ does not affect the number of offsets maintained by a process. However, it affects the size of each offset. Specifically, the max value of the offset is $\epsilon=\frac{\Bigepsilon}{\intervalsize}$ and the number of bits required for each offset is $\log_2 \epsilon$. Hence, a larger value of $\intervalsize$ is better for reducing the size of the $\newhvc$. However, with larger $\intervalsize$, the guarantees provided by $\newhvc$ are lower. Specifically, Lemma 3 shows that some unforced reordering may occur when events $e$ and $f$ differ by time $\Bigepsilon+\intervalsize$. Users should therefore choose the value of $\intervalsize$ based on the desired guarantees of $\newhvc$ or the maximum desirable offset. 

    
\begin{figure}[ht]
  \centering

  \begin{subfigure}{0.22\textwidth}
    \includegraphics[width=\linewidth]{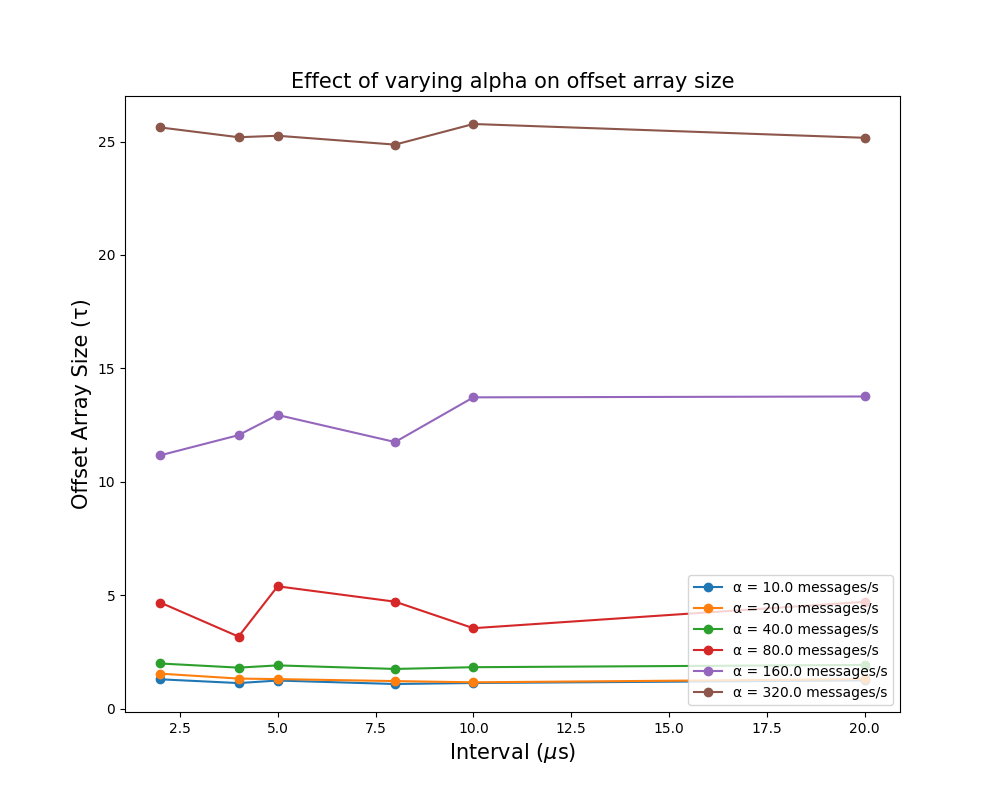}
    \caption{$n$ = 32.}
    \label{fig:effectofint-alpha1}
  \end{subfigure}
  \hfill
  \begin{subfigure}{0.22\textwidth}
    \includegraphics[width=\linewidth]{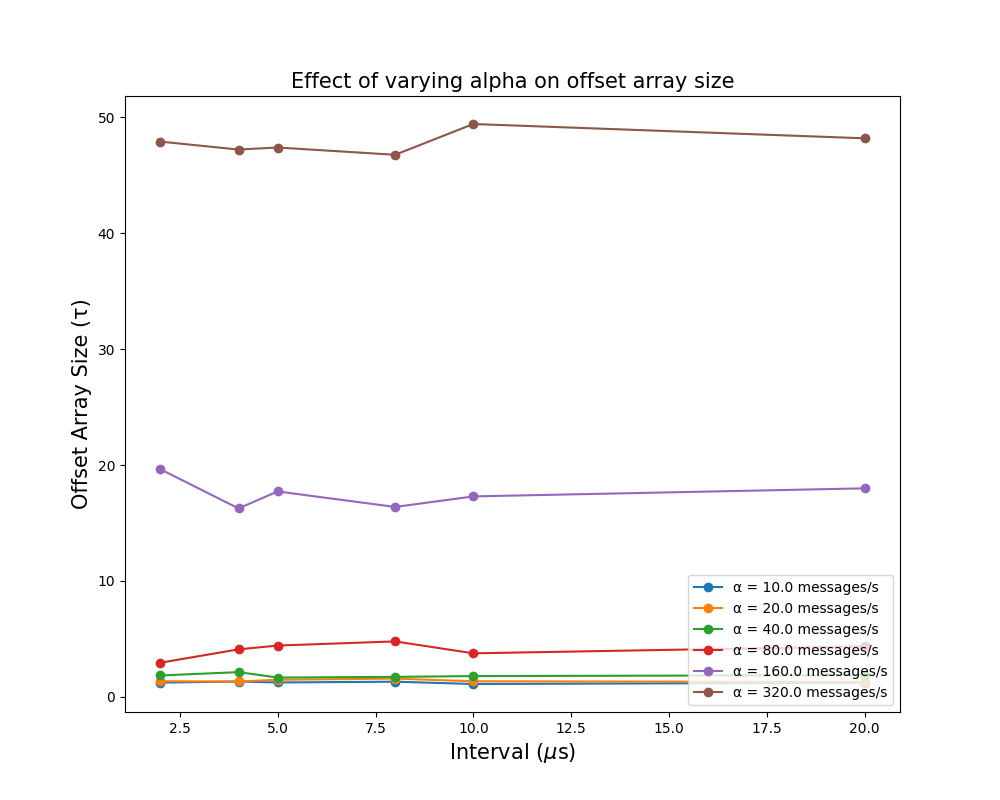}
    \caption{$n$ = 64.}
    \label{fig:effectofint-alpha2}
  \end{subfigure}

  \caption{$\offsetsize$ vs $\intervalsize$ when varying $\alpha$, $\Bigepsilon$ = 2 ms, $\delta$ = 8 $\mu s$.}
  \label{fig:effectofint-alphaoff}
\end{figure}

\textbf{Varying $\alpha$: }For every point in the $\intervalsize-\offsetsize$ trend, lower $\alpha$ values produce lower $\offsetsize$. This is consistent with our observations so far as communication is infrequent and processes tend to not hear from other processes, subsequently not storing their offsets. As $\intervalsize$ increases, we see this property respected, and $\offsetsize$ increasing consistently. This is illustrated in Figure \ref{fig:effectofint-alphaoff}.

\subsection{Effect of Message Delay (\texorpdfstring{$\delta$}{delta})}
\begin{figure}[ht]
  \centering

  \begin{subfigure}{0.22\textwidth}
    \includegraphics[width=\linewidth]{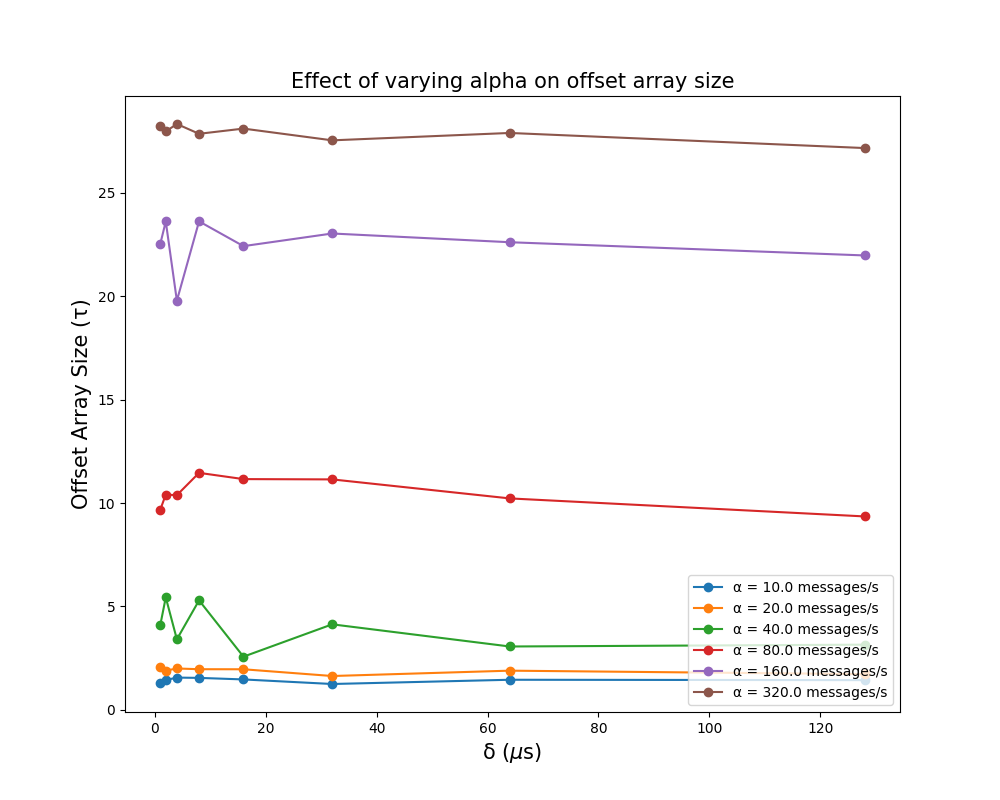}
    \caption{$n$ = 32.}
    \label{fig:effectofdelta-alpha1}
  \end{subfigure}
  \hfill
  \begin{subfigure}{0.22\textwidth}
    \includegraphics[width=\linewidth]{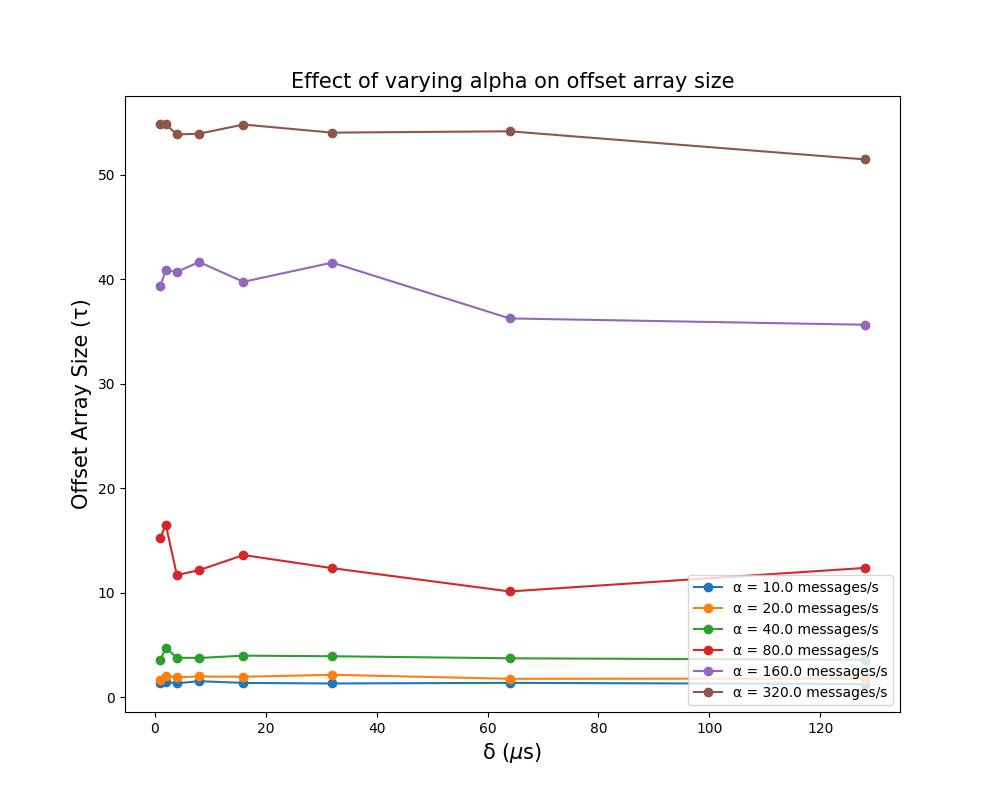}
    \caption{$n$ = 64.}
    \label{fig:effectofdelta-alpha2}
  \end{subfigure}

  \caption{$\offsetsize$ vs $\delta$ when varying $\alpha$, $\Bigepsilon$ = 4 ms, $\intervalsize$ = 16 $\mu s$.}
  \label{fig:effectofdelta-alphaoff}
\end{figure}

In this section, we observe the effect of $\delta$ on $\offsetsize$  while fixing $\Bigepsilon$ and $\intervalsize$. Here, we observe that the value of $\delta$ has minimal effect on $\offsetsize$. Specifically, as shown in Figure \ref{fig:effectofdelta-alphaoff}, we observe that the value of $\offsetsize$ increases as the value of $\alpha$ increases. However, for a fixed value of $\alpha$, the offset size remains the same. This is illustrated in Figure \ref{fig:effectofdelta-alphaoff}.




\subsection{Feasibility Regions}

In this section, we review the simulations to define the notion of feasible regions. As discussed earlier, the goal of $\newhvc$ is to enable the replay of a distributed computation with a small overhead. Here, we consider the case where the user identifies the expected overhead of $\newhvc$ to identify scenarios under which $\newhvc$ can be used to provide a \textit{perfect-replay} that meets all the requirements from Section \ref{sec:properties}. 
Since the overhead of the counters remains virtually unchanged, we only focus on the overhead of the number of offsets, i.e., the value of $\tau$. 

\begin{figure}[ht]
    \centering

    \begin{subfigure}[b]{0.22\textwidth}
        \centering
        \includegraphics[width=\textwidth]{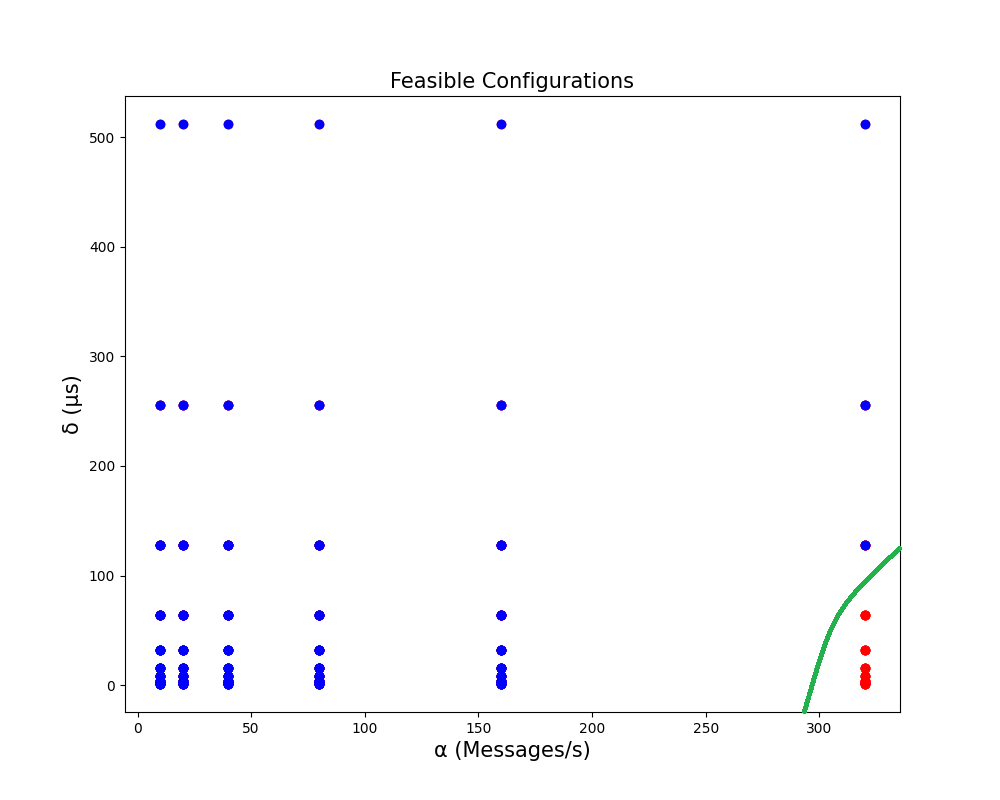}
        \caption{$\Bigepsilon$ = 1 ms, $n$ = 32, and $\offsetsize$ = 8.}
        \label{subfig:feasible.N32.E1000}
    \end{subfigure}
    \hfill
    \begin{subfigure}[b]{0.22\textwidth}
        \centering
        \includegraphics[width=\textwidth]{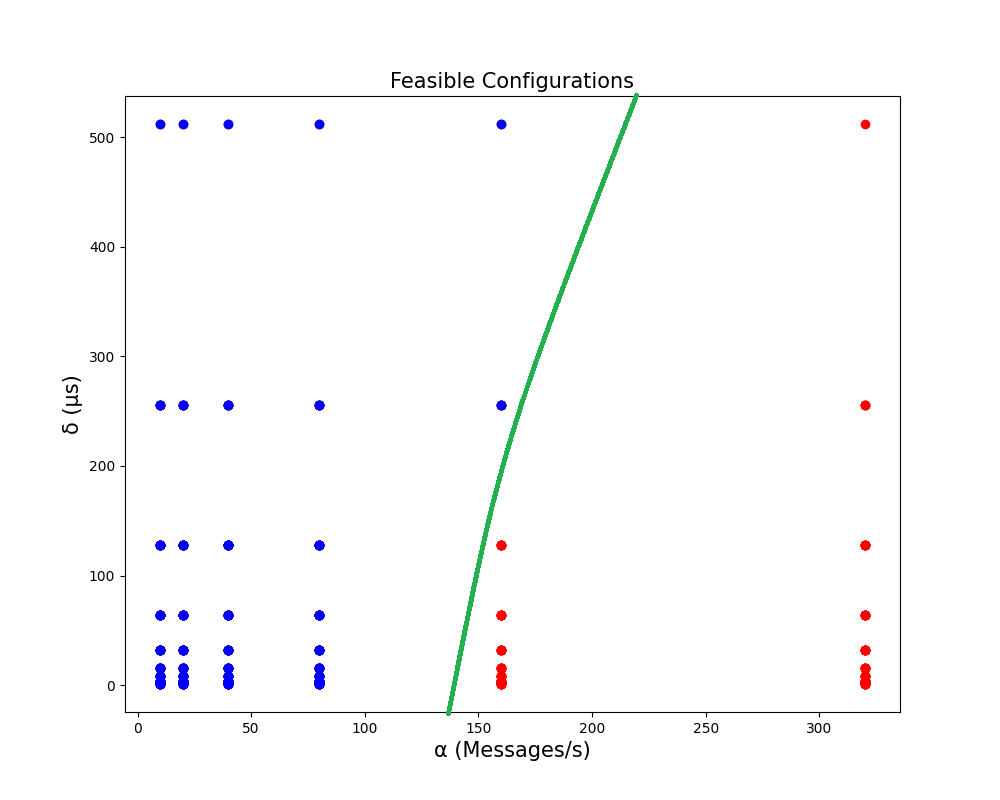}
        \caption{$\Bigepsilon$ = 2 ms, $n$ = 32, and $\offsetsize$ = 8.}
        \label{subfig:feasible.N32.E2000}
     \end{subfigure}
    \hfill
    \begin{subfigure}[b]{0.22\textwidth}
        \centering
        \includegraphics[width=\textwidth]{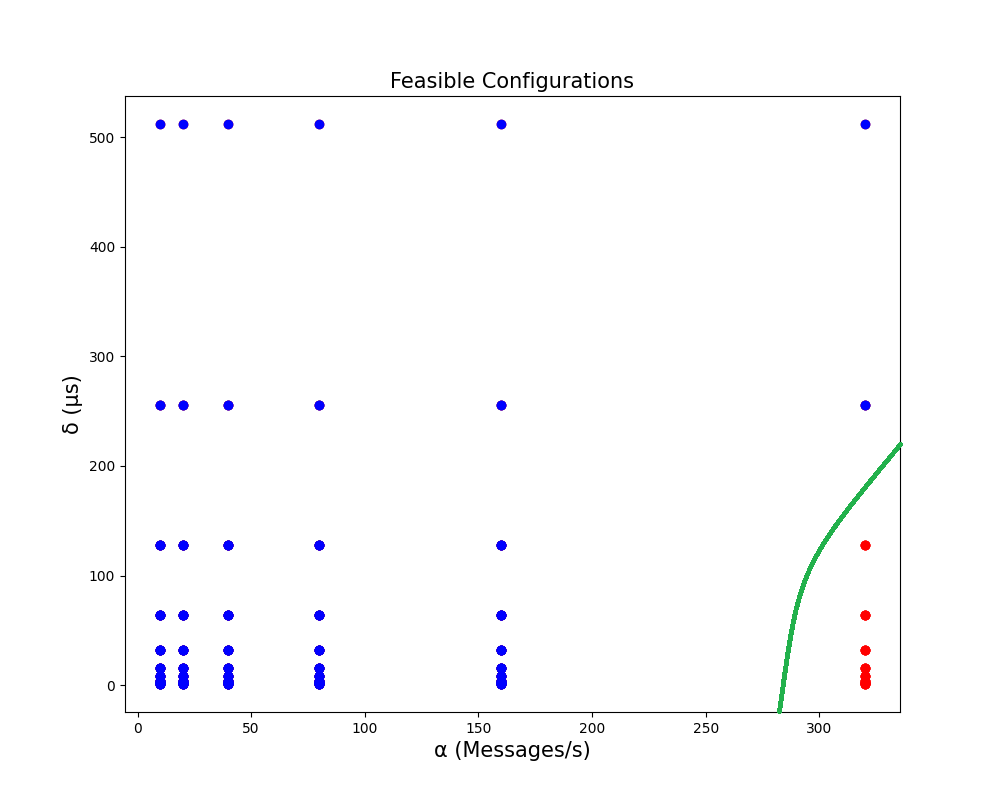}
        \caption{$\Bigepsilon$ = 1 ms, $n$ = 64, and $\offsetsize$ = 8.}
        \label{subfig:feasible.N64.E1000}
    \end{subfigure}
    \hfill
    \begin{subfigure}[b]{0.22\textwidth}
        \centering
        \includegraphics[width=\textwidth]{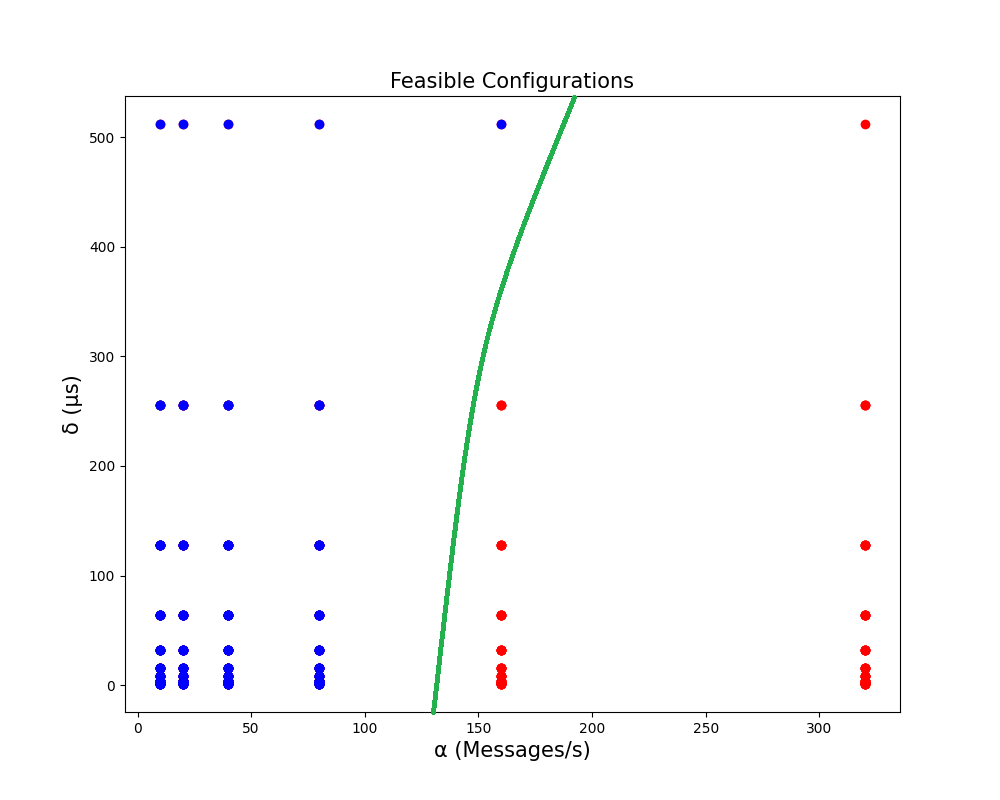}
        \caption{$\Bigepsilon$ = 2 ms, $n$ = 64, and $\offsetsize$ = 8.}
        \label{subfig:feasible.N64.E2000}
        
    \end{subfigure}

    \caption{Feasibility regions for alpha and delta settings.}
    \label{fig:feasibility-regions}
\end{figure}

For $\tau=8$, the feasibility regions are shown in Figure \ref{subfig:feasible.N32.E1000}. Here, the blue dots identify the data points where $\tau=8$ is feasible and the red dots represent the data points where $\tau=8$ is not feasible. The green line identifies the bounds where $\tau=8$ is feasible. We find that the size of the feasible region remains fairly unchanged with the value of $n$. However, it shrinks when the value of $\Bigepsilon$ is increased. This is expected based on how $\tau$ changes with $\Bigepsilon$. 

We note that the feasibility region only identifies the case where \textit{perfect-replay} meets all the requirements from Section \ref{sec:replay}. If the user needs to utilize $\newhvc$ in an infeasible region, the user can obtain \textit{partial-replay}. To understand this, consider the case where the actual value of $\Bigepsilon$ is $4ms$ but the user specifies it to be $2ms$ while constructing $\newhvc$. In this case, if $e$ and $f$ are within $2ms$ then $\newhvc$ will allow them to be replayed in any order. However, if $f$ occurred $3ms$ after $e$ then $e$ will always be replayed before $f$. We anticipate that even in a system where the clock skew is $4ms$, the actual clock skew at a given moment is likely to be smaller than $4ms$. This implies that the forced order between $e$ and $f$ will be quite infrequent. Hence, we anticipate that $\newhvc$ will be applicable even in domains where the system parameters cause it to fall in an infeasible region. 


\section{Related Work}
\label{sec:related}

The notion of a Logical clock (LC) was proposed in 1978 by Leslie Lamport \cite{Lamport78CACM} to trace the ordering of events in a distributed system. LCs do not use physical time (NTP \cite{RFC0958NTP}) and define a causal relation of $hb$ (happens-before), based on the communication between nodes (logical events). Vector time was designed independently by multiple researchers \cite{mattern1988virtual}\cite{Fidge87}\cite{DBLP:journals/ipl/SinghalK92}, and they proposed the idea of representing time in a distributed system as a set of n-dimensional non-negative integer vectors. Vector clock algorithms suggest that if two events $x$ and $y$ have timestamps $vh$ and $vk$, respectively, then $x$ happened before $y$then $vh < vk$. 

While vector clocks are upper bounded by $O(n)$ complexity in terms of both time and memory complexity, the different implementations of the past have tried to reduce this complexity and generate more efficient representations, with some success. Singhal-Kshemkalyani's differential technique \cite{DBLP:journals/ipl/SinghalK92}, relied on piggybacking using the last sent and last update, without updating every vector clock. This method relies on the assumption that even though the number of processes is large, only a few key processes in a system would interact frequently by passing messages. A benefit of this method is that it cuts down storage overhead at each process to $O(n)$. However, this method doesn't make a substantial contribution to reducing the time complexity incurred when updating the vector clock, as it relies on piggybacking to work. 
Fowler-Zwaenepoel's\cite{Fowler1990CausalDB} direct-dependency technique cuts down storage complexity again by reducing the message size during transmission by transmitting only a scalar value in the messages. The downside of this method is that it has a high computational overhead as it has to trace dependencies and update the vector clock, especially in systems where a few key processes may have a large number of events. 
Since all these approaches provide the same functionality as that of a vector clock, they will all have undesirable replays as identified in Section \ref{sec:replay}.

One existing limitation between vector clocks representing logical time and physical clock synchronization is the difficulty in reconciling one with the other. To overcome this challenge, Hybrid Logical clocks were introduced by Kulkarni et al. \cite{kulkarni2014logical} to capture the causality relationship of a logical clock with the characteristics of a physical clock embedded into it. Another variant of the hybrid clock is the Hybrid Vector Clock \cite{yingchareonthawornchai2018analysis}, which, unlike the Hybrid Logical Clock, is able to provide all possible/potential consistent snapshots for a given time. However, the HVC suffers from large overheads in execution, while updating all timestamps in the event of a send or receive. Our work with Replay Clocks expands on the idea of the Hybrid Vector Clock and provides an efficient implementation to counteract the large overheads of the HVC.

There has been some work in trace generation using event ordering, but most of these use the traditional logical/vector clock infrastructure. Dapper \cite{sigelman2010dapper} is Google's tracing software for distributed systems using logical event ordering. However, the authors mention that Dapper cannot correctly point to causal history, as it uses annotations in non-standard control primitives, that may mislead the causality calculations. ShiViz \cite{beschastnikh2016debugging} uses VCs in generating distributed system traces using happens-before relationships. However, since it uses traditional vector clocks, it uses a higher complexity than our proposed model.  

\section{Discussion}
\label{sec:discussion}





In Section \ref{sec:simulation}, we identified feasible regions for the given permissible overhead for $\newhvc$. Thus, the natural question is: \textit{what can a user do if the given system parameters fall into the infeasible region?} Here, observe that $\Bigepsilon$ provides one way to reduce the overhead if we accept some imperfect replay. To explain this, consider the case where we are using a system with clock skew to be $\Bigepsilon_a$. If the user implements $\newhvc$ with  $\Bigepsilon < \Bigepsilon_a$ then the resulting replay will still satisfy requirements 1 and 2 (cf. Section \ref{sec:newhvcrequirements}). Requirement 3 will be satisfied with $\epsilontwo=\Bigepsilon-\intervalsize$.

\begin{figure}
    \centering
    \includegraphics[width=0.5\linewidth]{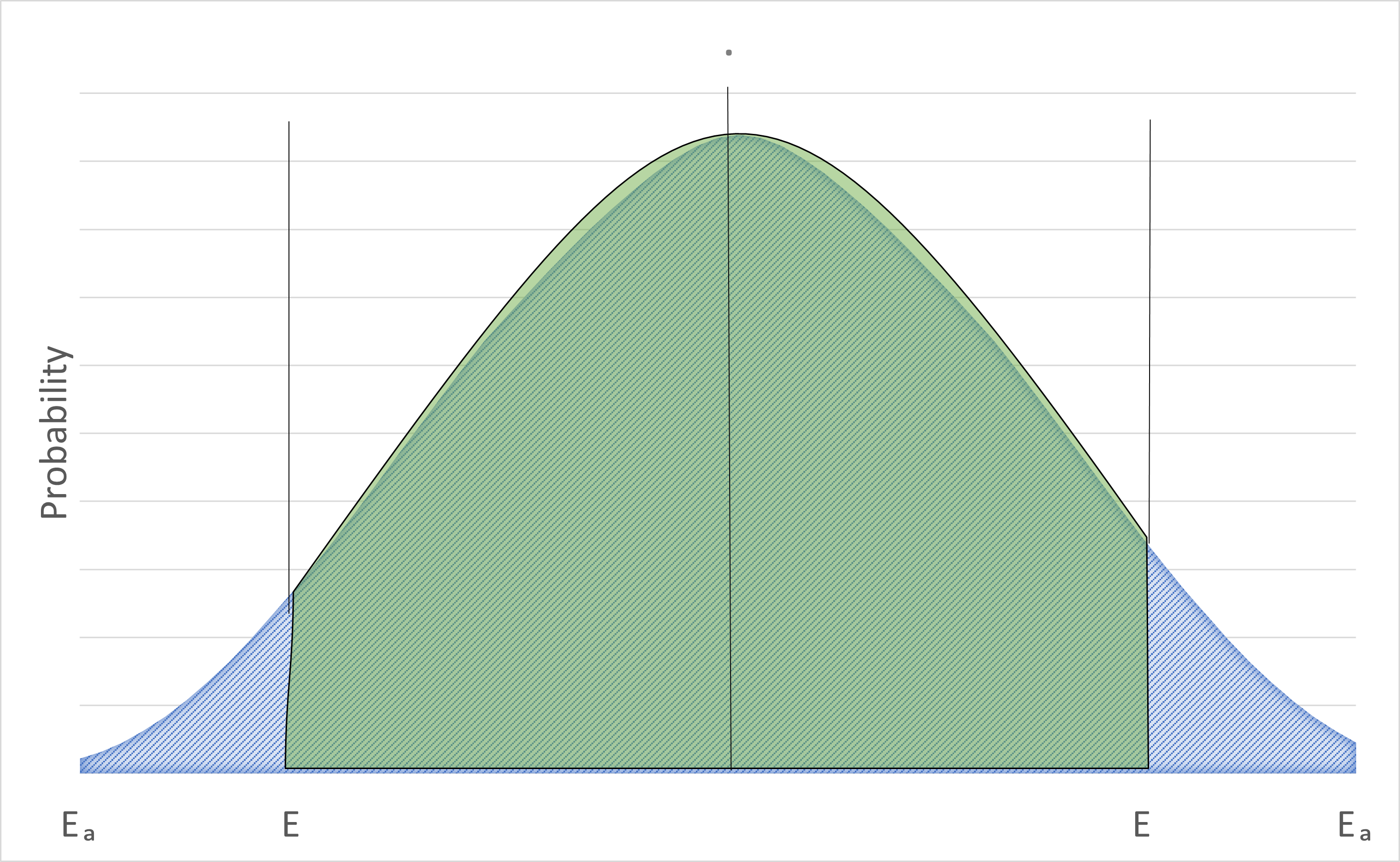}
    \caption{Effect of using $\Bigepsilon$ instead of $\Bigepsilon_a$ in $\newhvc$.}
    \label{fig:epsvsprob}
\end{figure}



Looking at this situation closely, we observe that the clock skew between two processes follows a structure shown in Figure \ref{fig:epsvsprob}. Specifically, at a given instance, the clocks of two processes $j$ and $k$ differ by some amount that is less than $\Bigepsilon_a$. However, the actual clock difference at a fixed point in time (that is not visible to either $j$ and $k$) is often less than $\Bigepsilon_a$. Hence, if $e$ and $f$ occurred at the same global time, the probability that the respective clocks differed by $\Bigepsilon$ depends upon the area of the shaded part (cf. Figure \ref{fig:epsvsprob}). In this case, $e$ and $f$ would still be replayed in arbitrary order. Only if the clocks fall in the non-shaded area then the replay will force an order between $e$ and $f$. In other words, even if the system parameters fall in the infeasible region, it would be possible to use $\newhvc$ that provides a valid replay. It is just that it will not be able to reproduce all possible replays.



\section{Conclusion and Future Work}
\label{sec:concl}

In this paper, we focused on the problem of replay clocks in systems where clocks are synchronized to be within $\Bigepsilon$. The purpose of these clocks is to reproduce a distributed computation with all its certainties and uncertainties. By certainty, we mean that if event $e$ must have happened before $f$ then the replay must ensure that $e$ is replayed before $f$. Specifically, this required that if $e$ happened before $f$ (as defined in \cite{Lamport78CACM}) or $f$ occurred $\epsilonone\approx\Bigepsilon$ time after $e$ then $e$ must occur before $f$.
And, by uncertainty, we mean that if $e$ and $f$ could occur in any order then the replay should not force an order between them. Specifically, if $e||f$ (as defined in \cite{Lamport78CACM}) and $e$ and $f$ occurred within time $\epsilontwo\approx\Bigepsilon$ then the replay permits them to be replayed in any order. 

We presented $\newhvc$ to solve the replay problem with $\epsilonone=\Bigepsilon+\intervalsize$ and $\epsilontwo=\Bigepsilon-\intervalsize$, where $\intervalsize$ is a parameter to $\newhvc$.  
We analyzed $\newhvc$ for various system parameters (clock skew ($\Bigepsilon$), message rate ($\alpha$), message delay ($\delta$)). We find that for various system parameters, the size of $\newhvc$ and the overhead to create timestamps and/or compare them is small.

For the purpose of replay, $\newhvc$ provides several advantages over existing approaches. For example, unlike logical clocks, they do not force certain unneeded event ordering. They have a significantly lower overhead compared to vector clocks. Also, they do not generate illegitimate replays that can occur with the user of vector clocks. 

The overhead of $\newhvc.j$ depends upon the number of processes that communicate with $j$ (directly or indirectly) in $\Bigepsilon$ window. This is different from the case in vector clocks where the overhead is always proportional to the number of processes in the system. 

We have utilized $\newhvc$ to enable users to visualize a distributed computation (cf. Appendix). The goal of the visualization is to allow the user to identify an event $f$ where a failure occurred. Then, they can use a replay of events just preceding $f$ to determine whether the error would go away. If it does, it would imply that it is a synchronization error. Likewise, a user can replay some portion of the computation. Since the replay of events may occur in a different order, it will help identify potential synchronization errors. 

$\newhvc$ is designed mainly for offline analysis, where the event data is stored during execution and analyzed at a later time. However, $\newhvc$ can be used for run-time monitoring/analysis as well if the data related to the timestamps is sent to a monitor, that monitor could analyze it for potential properties of interest if the analysis can be done \textit{quickly}. However, a key challenge in this context will be whether the run-time monitors can keep up with the execution of the system. 

There are several future directions for $\newhvc$. If the size of $\newhvc$ needed for perfect replay is too large, the user can reduce the size of $\newhvc$ by choosing a lower value of $\Bigepsilon$. In this case, the resulting replay will force some ordering between concurrent events. One of the future works is to identify the effect of reducing $\Bigepsilon$ in this manner. Another future work is to build a robust API for $\newhvc$ that will simplify its use in practice.

\bibliography{MasterReferences,refs}

\newpage

\appendix

\section{Application of \newhvc}

 In this section, we discuss an application of $\newhvc$ so as to visualize a given computation. 
This visualization is intended for the case where the user observes that there is an error in the program. Let $f$ denote the event where the error was observed.
Now, the user can utilize $\newhvc$ and visualize the events in the immediate \textit{past} of $f$. The user can also replay the events in the recent past of $f$ to determine if the error goes away. If the error is caused by a lack of synchronization, it will allow the user to identify the missing synchronization either via visualization or via identifying all possible replays that are possible in the given computation. A sample visualization trace is shown in Figure \ref{fig:visualization} in Appendix. It can also be replayed using the Algorithm in Section \ref{sec:replay}. This algorithm chooses one of the FrontLine events at random. Allowing the user to select these events provides the ability to the user to evaluate `what-if' scenarios to reduce the number of replays. 

\begin{figure*}
    \centering
    \includegraphics[width=0.5\linewidth]{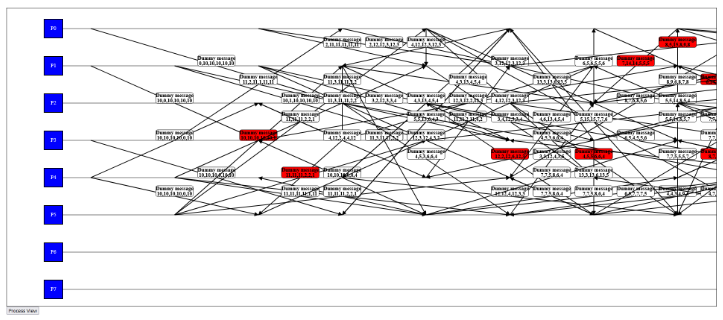}
    \caption{Sample API example of output traces with swimlanes in visualization.}
    \label{fig:visualization}
\end{figure*}

\section{Clock Size}

\begin{figure*}
  \centering

  \begin{subfigure}{0.3\textwidth}
    \includegraphics[width=\linewidth]{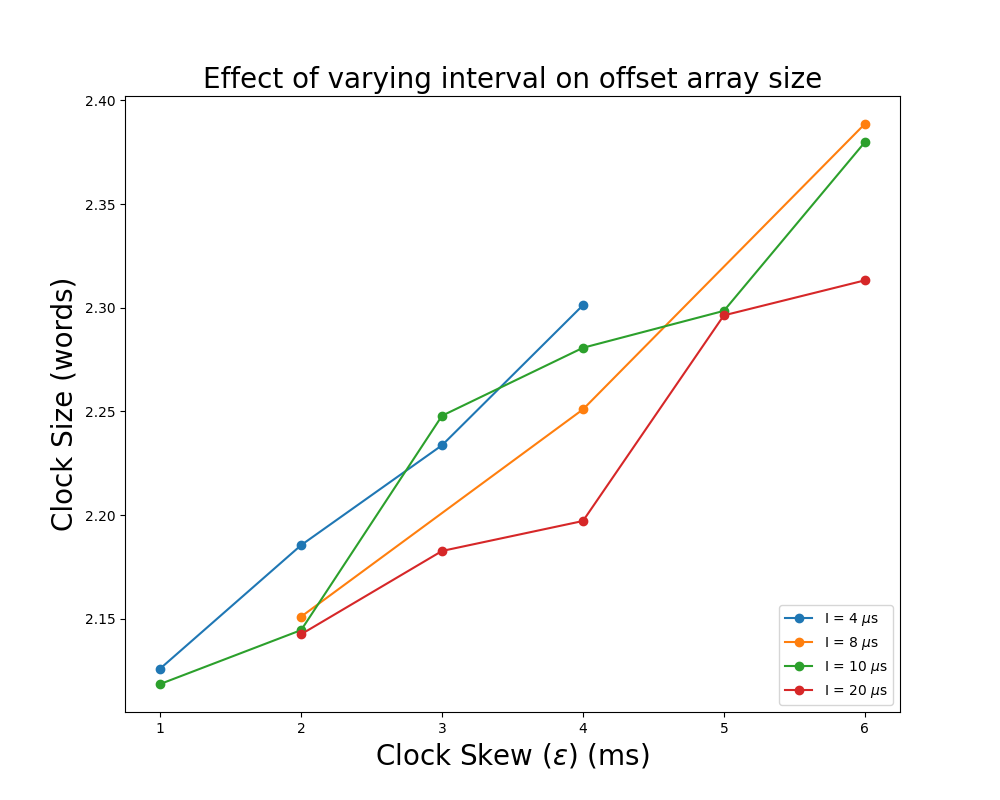}
    \caption{$\alpha$ = 20 messages/s, $n$ = 32.}
    \label{fig:effectofeps-interval1appendix}
  \end{subfigure}
  \hfill
  \begin{subfigure}{0.3\textwidth}
    \includegraphics[width=\linewidth]{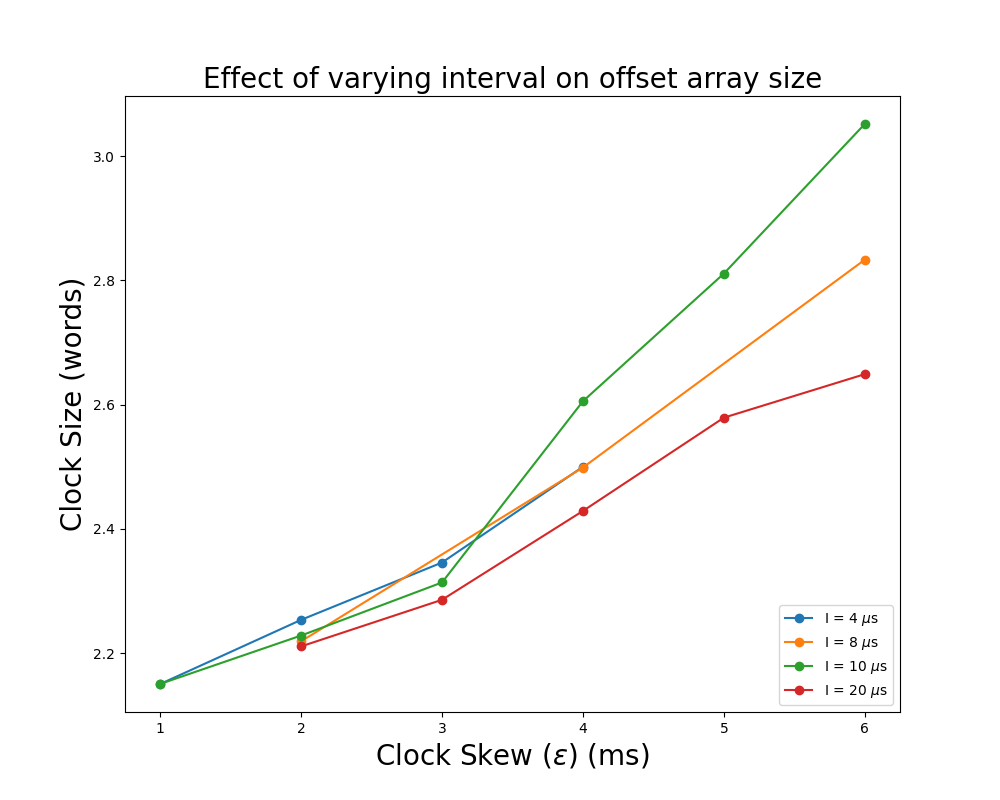}
    \caption{$\alpha$ = 40 messages/s, $n$ = 32.}
    \label{fig:effectofeps-interval2appendix}
  \end{subfigure}
  \hfill
  \begin{subfigure}{0.3\textwidth}
    \includegraphics[width=\linewidth]{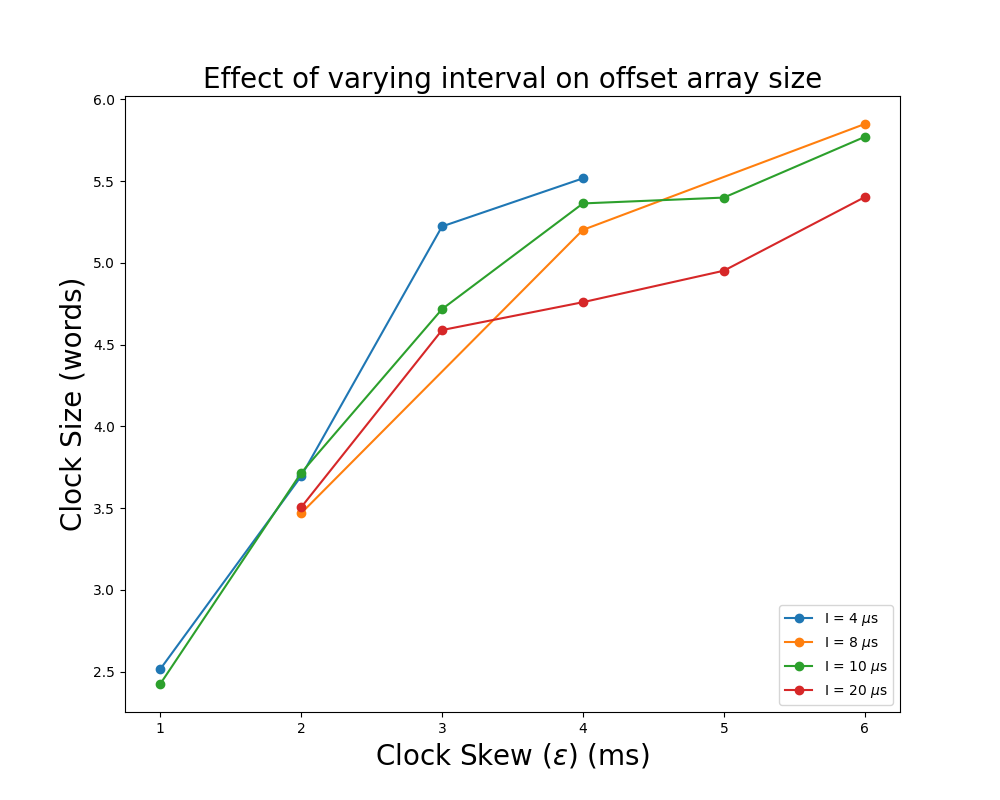}
    \caption{$\alpha$ = 160 messages/s, $n$ = 32.}
    \label{fig:effectofeps-interval3appendix}
  \end{subfigure}

  \begin{subfigure}{0.3\textwidth}
    \includegraphics[width=\linewidth]{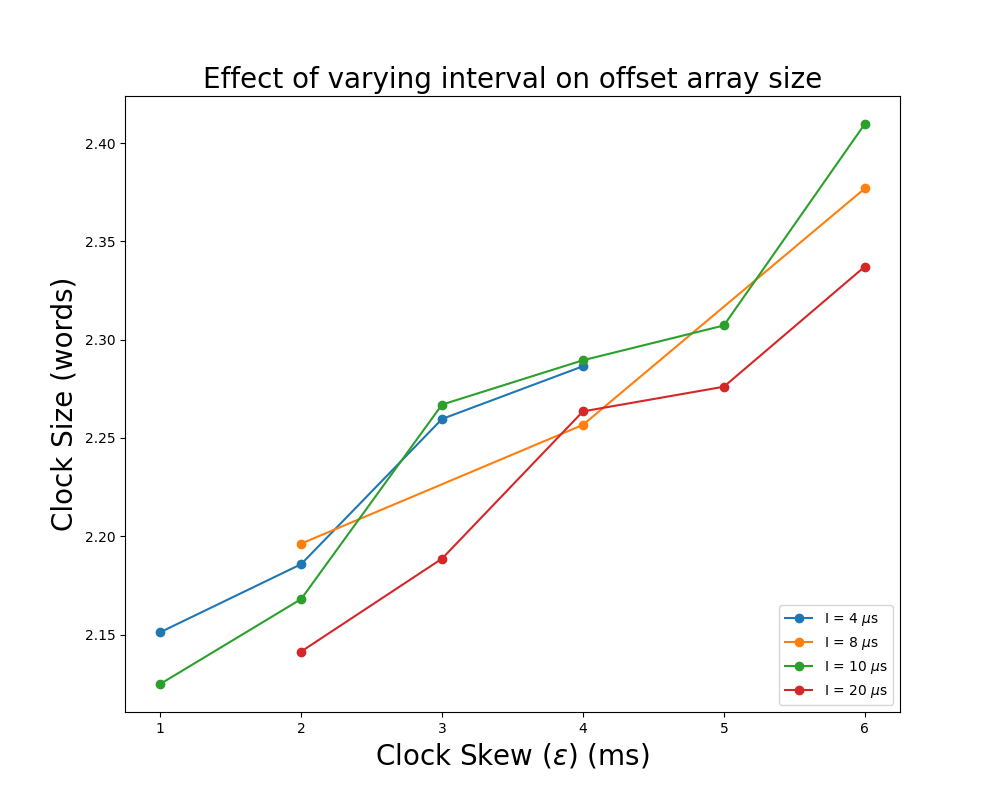}
    \caption{$\alpha$ = 20 messages/s, $n$ = 64.}
    \label{fig:effectofeps-interval4appendix}
  \end{subfigure}
  \hfill
  \begin{subfigure}{0.3\textwidth}
    \includegraphics[width=\linewidth]{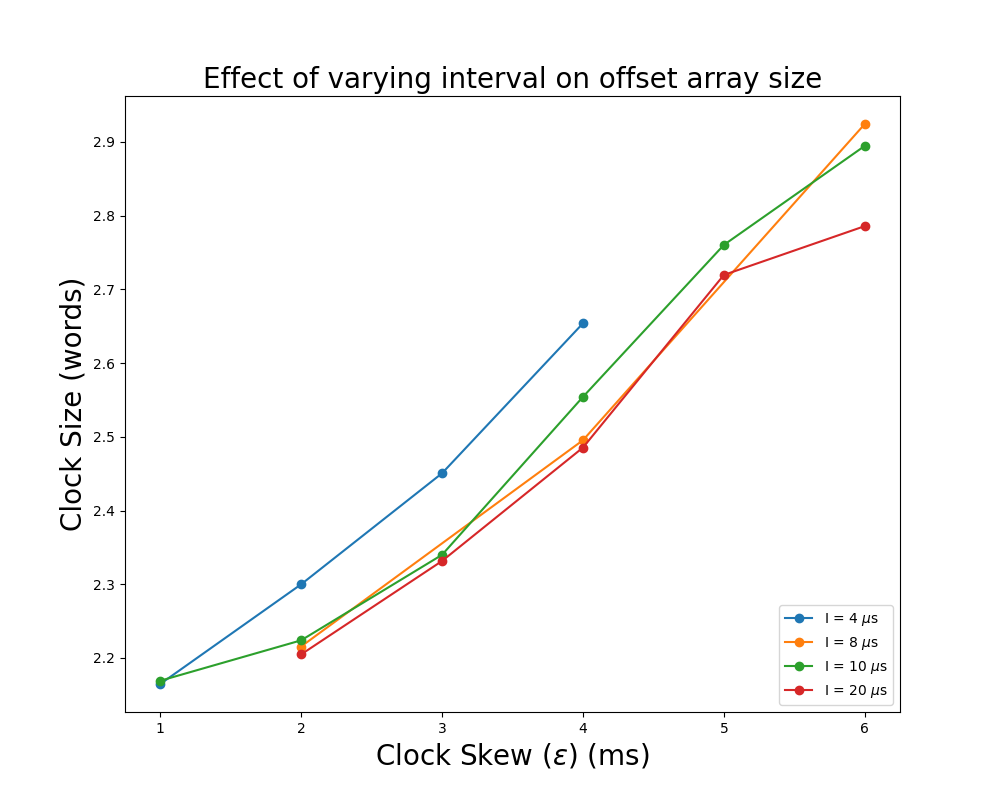}
    \caption{$\alpha$ = 40 messages/s, $n$= 64.}
    \label{fig:effectofeps-interval5appendix}
  \end{subfigure}
  \hfill
  \begin{subfigure}{0.3\textwidth}
    \includegraphics[width=\linewidth]{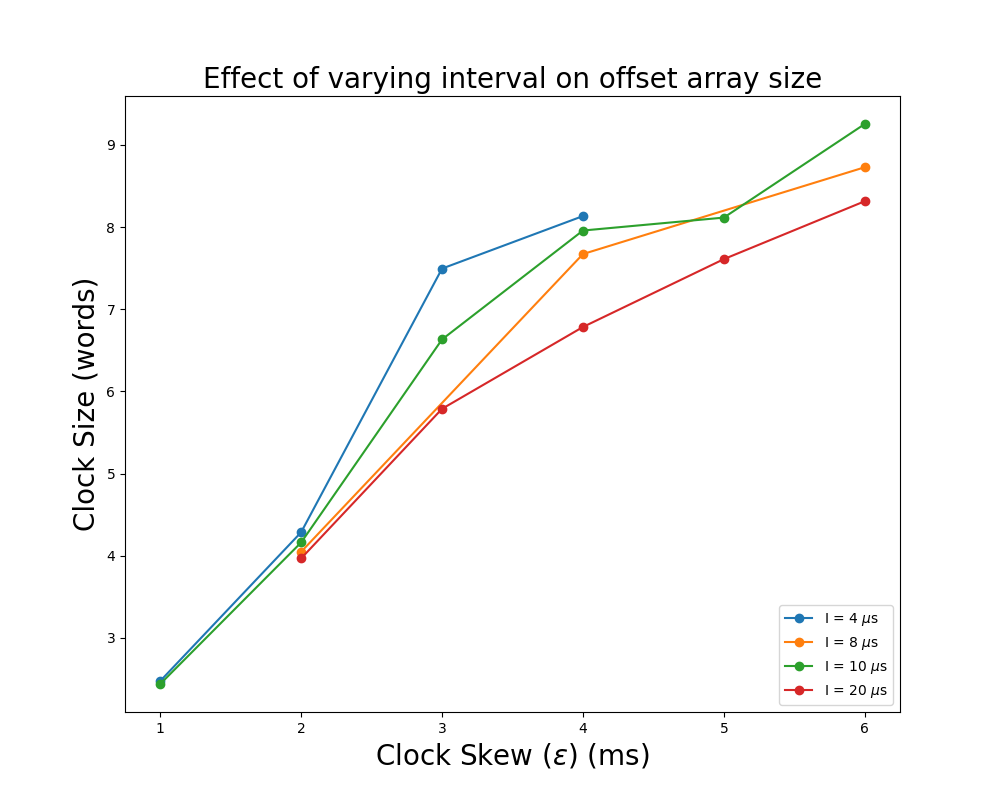}
    \caption{$\alpha$ = 160 messages/s, $n$ = 64.}
    \label{fig:effectofeps-interval6appendix}
  \end{subfigure}

  \caption{Clock Size vs $\Bigepsilon$ when varying $\intervalsize$, $\delta = 8 \mu s$.}
  \label{fig:effectofeps-intervaloffsetappendix}
\end{figure*}

Following from Section \ref{sec:simulation}, we present Figure \ref{fig:effectofeps-intervaloffsetappendix} as an extension to Figure \ref{fig:effectofeps-intervaloffset}. Here, we replace the $Y$-axis of $\offsetsize$ with the actual size of the clock in words. This relation is a linear transformation and produces the same trends observed in Figure \ref{fig:effectofeps-intervaloffset}. Following our discussion about the size of the clock in Section \ref{sec:representation}, the clock size for most cases is between 2 to 4 words for our most efficient configurations. This would reduce even further with choices to represent the sum of counters in the extraneous bits of the epoch store, as we do not lose as much information in most cases.

\vspace{-5em}

\end{document}